\documentclass[12pt,preprint]{aastex}

\shorttitle{Widespread hot plasma in non-flaring active region}
\shortauthors{Reale et al.}

\usepackage{times}

\begin{document}

\title{Evidence of widespread hot plasma in a non-flaring coronal active region from Hinode/XRT}

\author{Fabio Reale\altaffilmark{1}}
\affil{Dipartimento di Scienze Fisiche \& Astronomiche, Universit\`a di
       Palermo, Sezione di Astronomia, Piazza del Parlamento 1, 90134 Palermo,
       Italy}
\author{Paola Testa}
\affil{Harvard-Smithsonian Center for Astrophysics, Cambridge, MA 02138, USA}
\author{James A. Klimchuk}
\affil{NASA Goddard Space Flight Center, Greenbelt, MD 20771, USA}
\author{Susanna Parenti}
\affil{Royal Observatory of Belgium, 3 Circular Avenue, B-1180 Brussels, Belgium}

\altaffiltext{1}{INAF - Osservatorio Astronomico di Palermo ``G.S.
       Vaiana'', Piazza del Parlamento 1, 90134 Palermo, Italy}

\begin{abstract}
Nanoflares, short and intense heat pulses within spatially unresolved
magnetic strands, are now considered a leading candidate to solve the
coronal heating problem. However, the frequent occurrence of nanoflares
requires that flare-hot plasma be present in the corona at all times.
Its detection has proved elusive until now, in part because the
intensities are predicted to be very faint.  Here we report on the
analysis of an active region observed with five filters by Hinode/XRT in
November 2006. We have used the filter ratio method to derive maps of
temperature and emission measure both in soft and hard ratios.  These
maps are approximate in that the plasma is assumed to be isothermal
along each line-of-sight.  Nonetheless, the hardest available ratio
reveals the clear presence of plasma around 10 MK.  To obtain more
detailed information about the plasma properties, we have performed
Monte Carlo simulations assuming a variety of non-isothermal emission
measure distributions along the lines-of-sight.  We find that the
observed filter ratios imply bi-modal distributions consisting of a
strong cool ($\log T \sim 6.3-6.5$) component and a weaker (few percent)
and hotter ($6.6 < \log T < 7.2$) component.  The data are consistent
with bi-modal distributions along all lines of sight, i.e., throughout
the active region.  We also find that the isothermal temperature
inferred from a filter ratio depends sensitively on the precise
temperature of the cool component.  A slight shift of this component can
cause the hot component to be obscured in a hard ratio measurement.
Consequently, temperature maps made in hard and soft ratios tend to be
anti-correlated.  We conclude that this observation supports the
presence of widespread nanoflaring activity in the active region.
\end{abstract}


\section{Introduction}

The solar corona, once thought to be heated in a quasi-steady
fashion, has proven to be much more difficult to explain than early
observations suggested.  Developements over the last few years point
to a different picture in which coronal loops and perhaps also the
diffuse corona are heated impulsively by nanoflares occurring within
unresolved strands
\citep{1988ApJ...330..474P,1994ApJ...422..381C,1997ApJ...478..799C,
2006SoPh..234...41K,2003ApJ...593.1174W,2006ApJ...651.1219P}.
Nanoflares can explain a number of observations that are clearly
inconsistent with static models, such as the density excess and
superhydrostatic scale height of $\sim 1$ MK loops.  Furthermore,
the proposed mechanisms for coronal heating predict that the heating
should be impulsive when viewed properly from the perspective of
individual magnetic flux strands \citep{2006SoPh..234...41K}. Despite
these successes, the existence of nanoflares is still under debate.
No ``smoking gun" has yet been found.

The strongest evidence in support of nanoflares would be the
detection of very hot plasma \citep{1995itsa.conf...17C}. 
Hydrodynamic
simulations show that impulsively heated strands should reach
temperatures in excess of 5 MK.  Much higher temperatures are
possible depending on the energy of the nanoflare. Until now, there
has been little evidence of such very hot plasma outside of proper
flares.  There are likely two reasons for this. First, few studies
have explicitly looked for such extreme temperatures during
quiescent (nonflaring) conditions. Second, and more importantly, the
intensity of the hot emission is expected to be very weak. The
reason is as follows. Temperature rises abruptly when a nanoflare
occurs, but density is much slower to respond. By the time
chromospheric evaporation has raised the density significantly, the
coronal plasma has cooled to temperatures far below the peak value.
Low densities combined with rapid initial cooling means that the hot
phase contributes relatively little to the total emission measure of
a heating and cooling event. Hydrodynamic simulations predict that
the emission measure of the hottest plasma is 2-3 orders less than
that of the dominant coronal plasma \citep{2008ApJ...682.1351K}.  The
emitted radiation is correspondingly faint. Ionization
nonequilibrium effects may diminish the hot component even more
\citep{2008ApJ...684..715R,2006A&A...458..987B}.

Recent observations suggest that very hot plasma may indeed be
present at low levels in nonflaring active regions
\citep{2006SoSyR..40..272Z,2007AstL...33..396U,patsou08,2008AGUFMSH13A1511M,2009arXiv0901.3122S}. We have therefore undertaken a focused investigation to determine as
precisely as possible the amount of very hot plasma in one
particular active region and to evaluate whether it is consistent
with nanoflare heating.  We use wide-band multi-filter imaging
observations \citep{2007Sci...318.1582R} from the X-Ray Telescope
(XRT, \cite{2007SoPh..243...63G}) on board the Hinode mission
(\cite{2007SoPh..243....3K}).  These observations have the advantage
of a strong signal compare to spectrometer observations of
individual spectral lines.  The disadvantage is that each band is
sensitive to a range of plasma temperatures.  To distinguish hot
from cool plasmas, we use different filter combinations as described below.

The major problem is to screen out the dominant cooler emitting
components which may inhibit the detection of the minor hot
component.  Hinode/XRT is equipped with several wide-band filters,
which might fulfill this task.  The filters with a band ranging down
to relatively low energies do not allow the detection of the hot
plasma, because it is overwhelmed by the much larger quantities of
cooler plasma. At the other extreme, flare filters hardly detect
flare-hot plasma with very low emission measure. 
There is a third opportunity offered by the wide choice of filters:
filters of intermediate thickness which may be the right compromise
between sensitivity level and energy range.

\section{The observation and preliminary analysis}

Through its nine filters in the soft X-ray band the XRT is sensitive
to the emission of plasma in the temperature range $6.1 <
\log~T<7.5$.  The CCD camera has a 1 arcsec pixel and the
Full-Width-at-Half-Maximum (FWHM) of the Point Spread Function (PSF)
is $\approx 0.8$ arcsec \citep{2007SoPh..243...63G}.  We consider the same data as those
already illustrated in \cite{2007Sci...318.1582R}. The
512$\times$512 arcsec$^2$ field of view includes active region
AR10923 observed close to the Sun center on 12 November 2006.  The
filters used were Al\_poly (F1), C\_poly (F2), Be\_thin (F3),
Be\_med (F4) and Al\_med (F5), with exposure times of 0.26~s,
0.36~s, 1.44~s, 8.19~s and 16.38~s, respectively.  The selected
dataset covers one hour, starting at 13:00 UT, and the time interval
between one exposure and the next in the same filter is about five
minutes (12 images in each filter). This active region was monitored
by the XRT all along its passage from the east to the west limb. It
flared several times during this period. However, during the
selected hour no major flare and no significant rearrangement of the
region morphology occurred (the average of the RMS intensity
deviation of the individual pixels is $\approx 12$\%), so that we
can average over the whole hour.  We used the up-to-date standard
XRT software to preprocess the data, including corrections for the
read-out signal, flat-field, CCD bias. The observed images were
co-aligned with a cross-correlation technique.


Each XRT filter has a different response to plasma temperature,
described by a function $G_{Fj}(T)$, where $j$ indicates the filter order
number. Thinner filters F1 and F2 are more sensitive to the emission
of cooler plasma than thicker filters F4 and F5.  Images of the same
region in different filters provide information about the temperature of
the emitting plasma. If the emitting plasma is isothermal and optically
thin along the line of sight, the ratio between signals in thick and thin
filters is a function of the temperature only. \cite{2007Sci...318.1582R}
devised a special filter ratio from a combination of all available
filters (Combined Improved Filter Ratio, CIFR), which optimizes for the
signal-to-noise ratio and reveals a very fine thermal structure of the
active region.

\begin{figure}[!t]
  \centering \includegraphics[width=8cm]{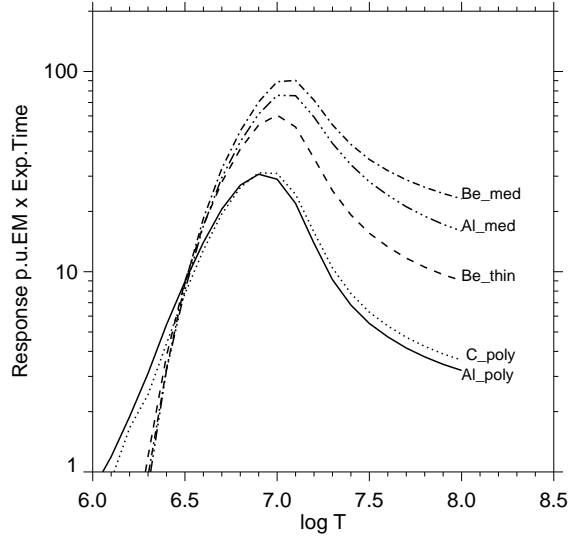} 
  \caption{Temperature response function of the 5 filters used in the XRT observation. To compensate for the different exposure times, the responses are multiplied by the respective total exposure times. 
 }
  \label{fig_rsp}
\end{figure}

The analysis presented here considers also several standard ratios of
single filters, and we take an up-to-date filter temperature calibration
with a special fine tuning to make the analysis entirely self-consistent
(see Appendix). Fig.~\ref{fig_rsp} shows the temperature response functions of the filters. Five available filters allow to define 10 different
ratios of single filters. We will exclude the ratio F2/F1 from our
analysis because the dependence on temperature is double-valued over
a large range.  We are then left with 9 relevant ratios. The 6 ratios
with F2 and F1 in the denominator, and also CIFR, are able to diagnose
plasmas down to $\log T \approx 6$, i.e. relatively cool plasma; the two
ratios with F3 in the denominator down to $\log T \approx 6.2$; and the
ratio F4/F5 down to $\log T \approx 6.3-6.4$.  An important point is that
the coolest plasma detectable by the last ratio is about twice as hot
as the coolest plasma detectable by the first two ratios.  The F4/F5
ratio is a monotonic function of $\log T$ in the range $6.4 \leq \log
T \leq 7.5$.  We have set minimum acceptance thresholds to compute the
ratios. All pixels with signal below 200 DN/s for F1, 100 DN/s for F2,
30 DN/s for F3, 2 DN/s for F4 and 10 DN/s for CIFR have been masked out.


\section{Analysis and results}

\subsection{Data analysis}

\begin{figure*}[!t]
  \centering \includegraphics[width=8cm]{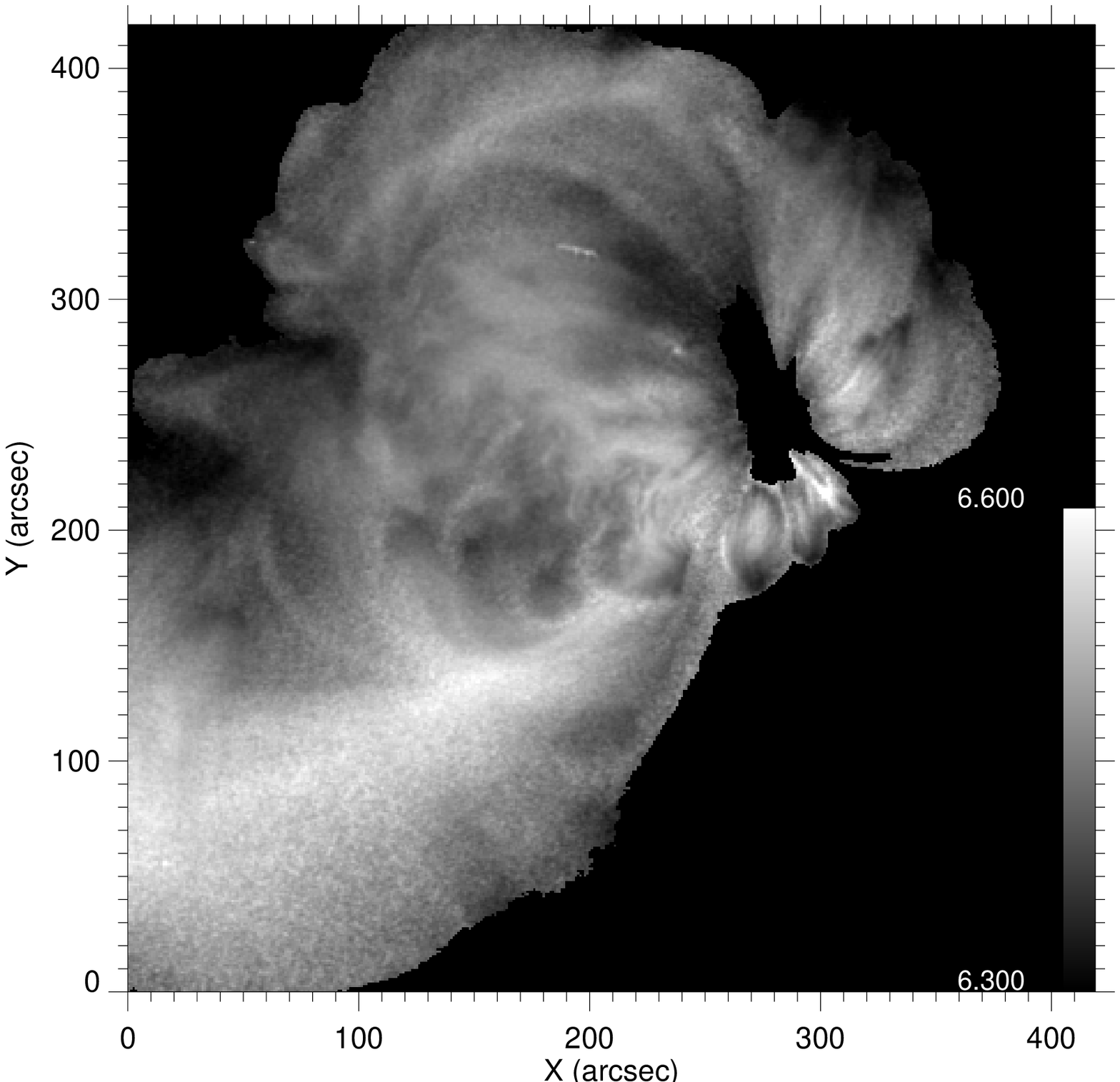} \includegraphics[width=8cm]{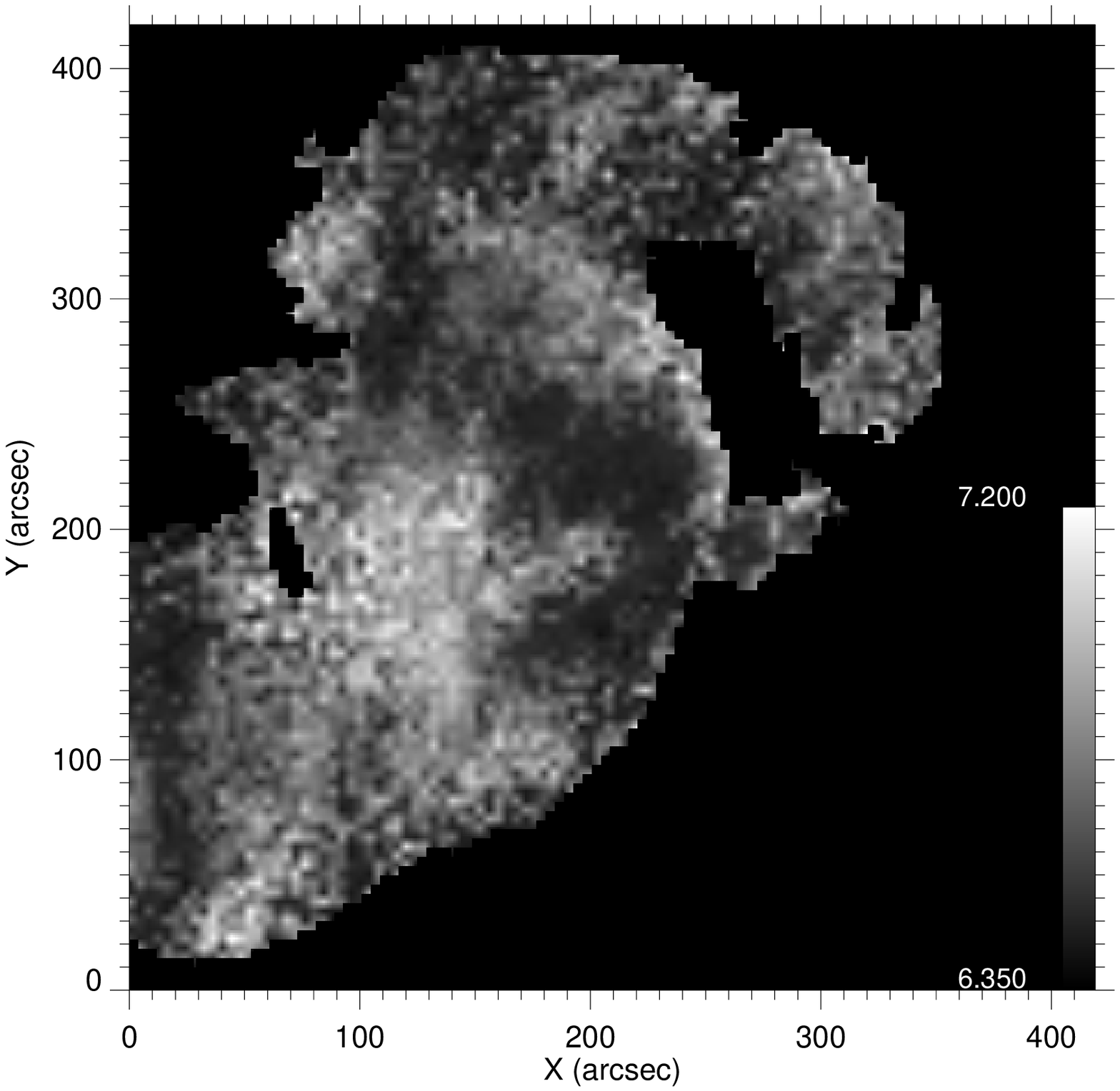}
  \centering \includegraphics[width=8cm]{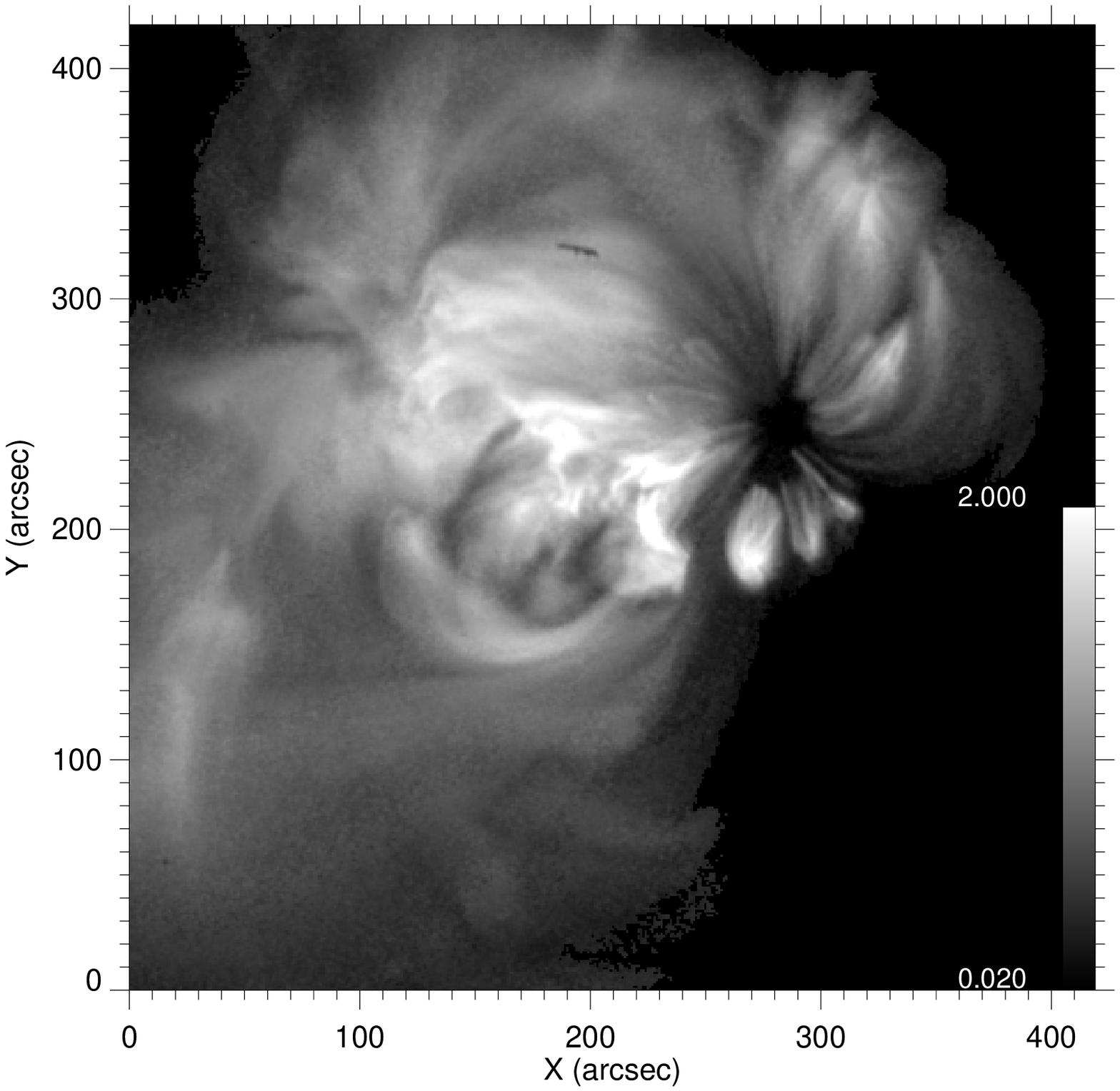} \includegraphics[width=8cm]{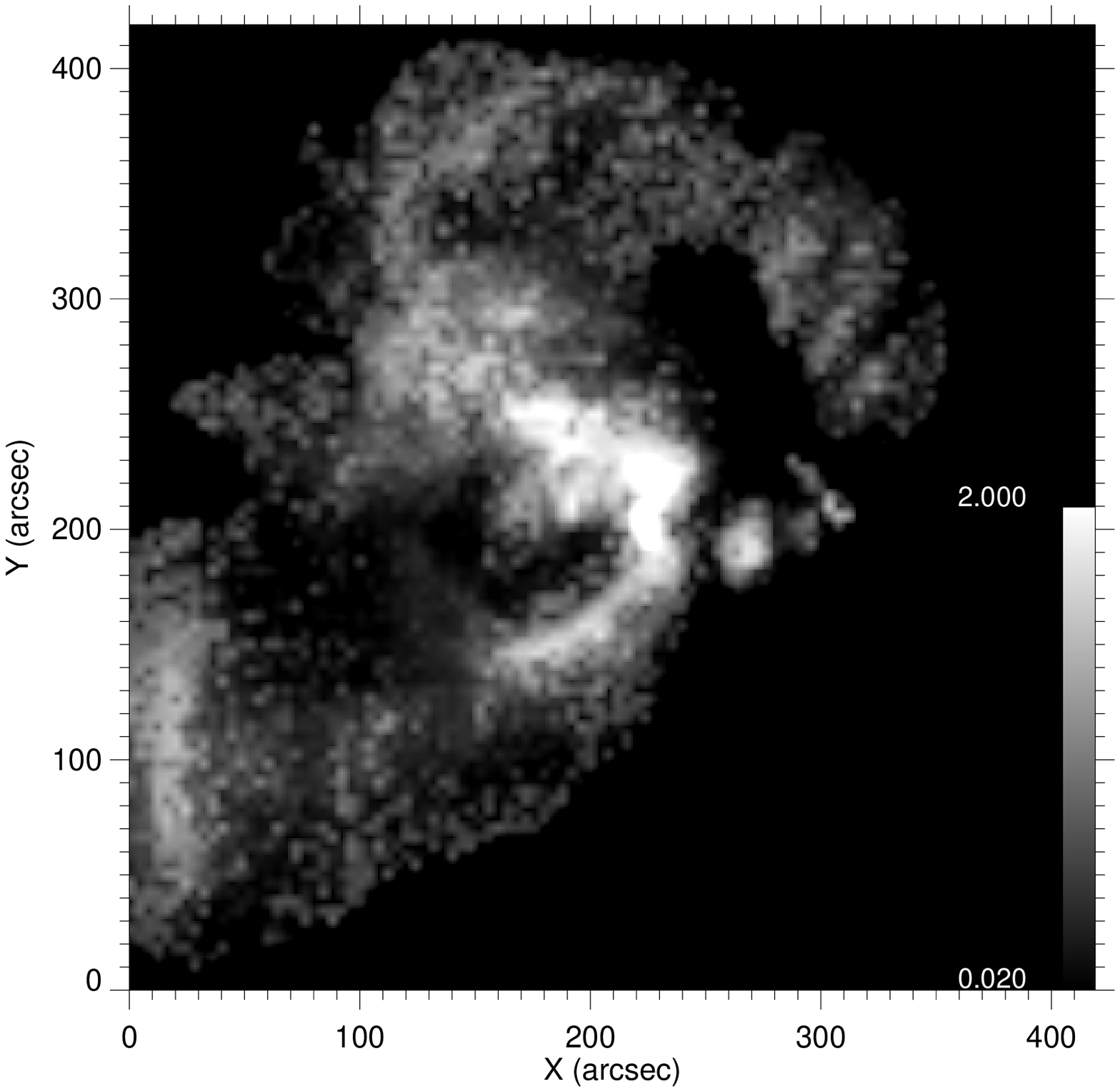}
  \caption{Maps of temperature (log T, upper row) and emission measure (lower row) obtained with CIFR (left column) and hard (F4/F5) filter ratio (right column). The emission measure is in units of $10^{44}$ cm$^{-3}$ s$^{-1}$. The F4/F5 maps are obtained after binning over boxes of 4$\times$4 pixels.
 }
  \label{fig1}
\end{figure*}

Fig.~\ref{fig1} shows temperature and emission measure maps obtained
with CIFR (left column) and with the hard (F4/F5) filter ratio. To
improve the signal-to-noise ratio, the F4/F5 maps are obtained after
binning over boxes of 4$\times$4 pixels.
We have estimated the uncertainties on temperature and emission measure in each pixel from the photon counts derived from the DN in each filter, with a conversion factor obtained assuming an average spectrum at central temperature about $\log T = 6.7$ (the conversion factor does not change much in the range $6.5 \leq \log T \leq 7$). Error propagation to temperature and emission measure has been applied according to \cite{1995ApJ...448..925K}. The uncertainties are globally consistent with the average root mean square fluctuations of the temperature and emission measure values of the pixels surrounding a given pixel \citep{2006A&A...449.1177R}. The average errors in temperature and emission measure obtained from CIFR in each pixel are around 0.004 ($\sim 1$\%) and 0.015 in the log, respectively, in regions of high signal (e.g. region SH in Fig.\ref{fig:mapsub}), and 0.008 and 0.03 in regions of low signal (e.g. region HH in Fig.\ref{fig:mapsub}). The errors from F4/F5 in each 4$\times$4 pixel box are around 0.05 and 0.2, respectively, in regions of high signal, and 0.1 and 0.25 in regions of low signal.

The difference in resolution and detail definition is immediately apparent: the softer
ratio maps show much better-defined structuring.  While the soft
ratio detects plasma mostly in the range $6.3 < \log T < 6.6$, the
hard ratio appears to detect hotter components to $\log T >7$.  In
particular we find hot plasma at the boundary of dense structures
and in an extensive region on the center left of the active region,
where the emission measure is instead low.  We also notice that
the temperature maps do not look similar, while the
emission measure maps to some extent do, e.g. we find high emission measure in the same zones of the active region.

Therefore we find different thermal conditions using different
filter ratios.  In particular, relatively hot regions in the CIFR
and other soft filter ratios appear as relatively cool in the hard
filter ratio. This is not trivially expected and deserves further
investigation, that we will show later.

Our interest is also to exploit the multi-band imaging observation
to obtain quantitative information about the thermal distribution of
the plasma both along the line of sight and across the active
region.

\begin{figure}[!t]
  \centering \includegraphics[width=12cm]{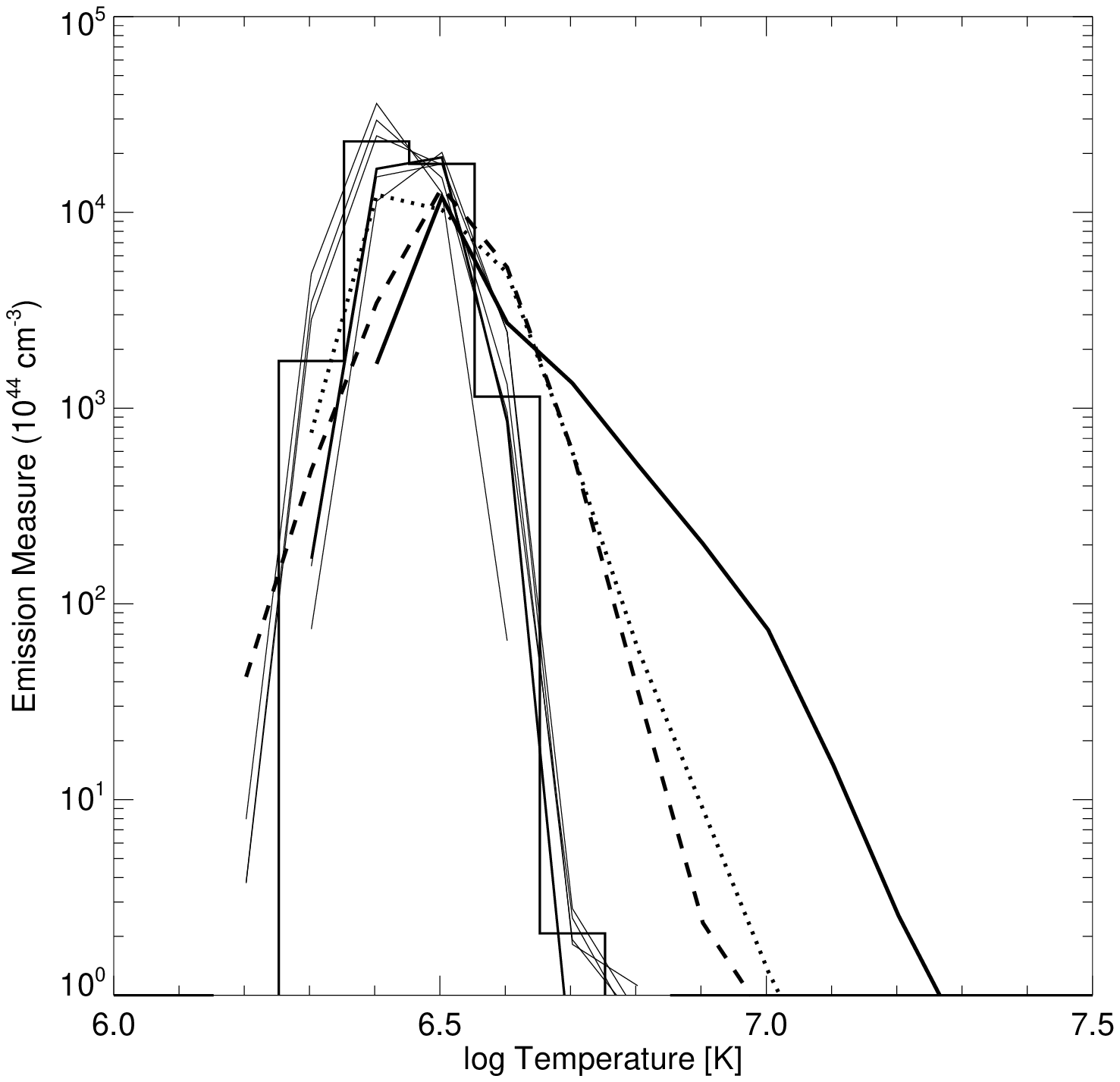}
  \caption{Emission measure distribution vs temperature for the whole
  active region (with emission above threshold of acceptance) obtained
  from available filter ratios: soft filter ratios (thin solid lines),
  CIFR (histogram), F5/F3 (dotted line), F4/F3 (dashed line), F4/F5
  (thick solid line). The latter is obtained from the map binned
  over
  boxes of 4$\times$4 pixels.
 }
  \label{fig:emt_ar}
\end{figure}

\begin{figure*}[!t]
  \centering \includegraphics[width=12cm]{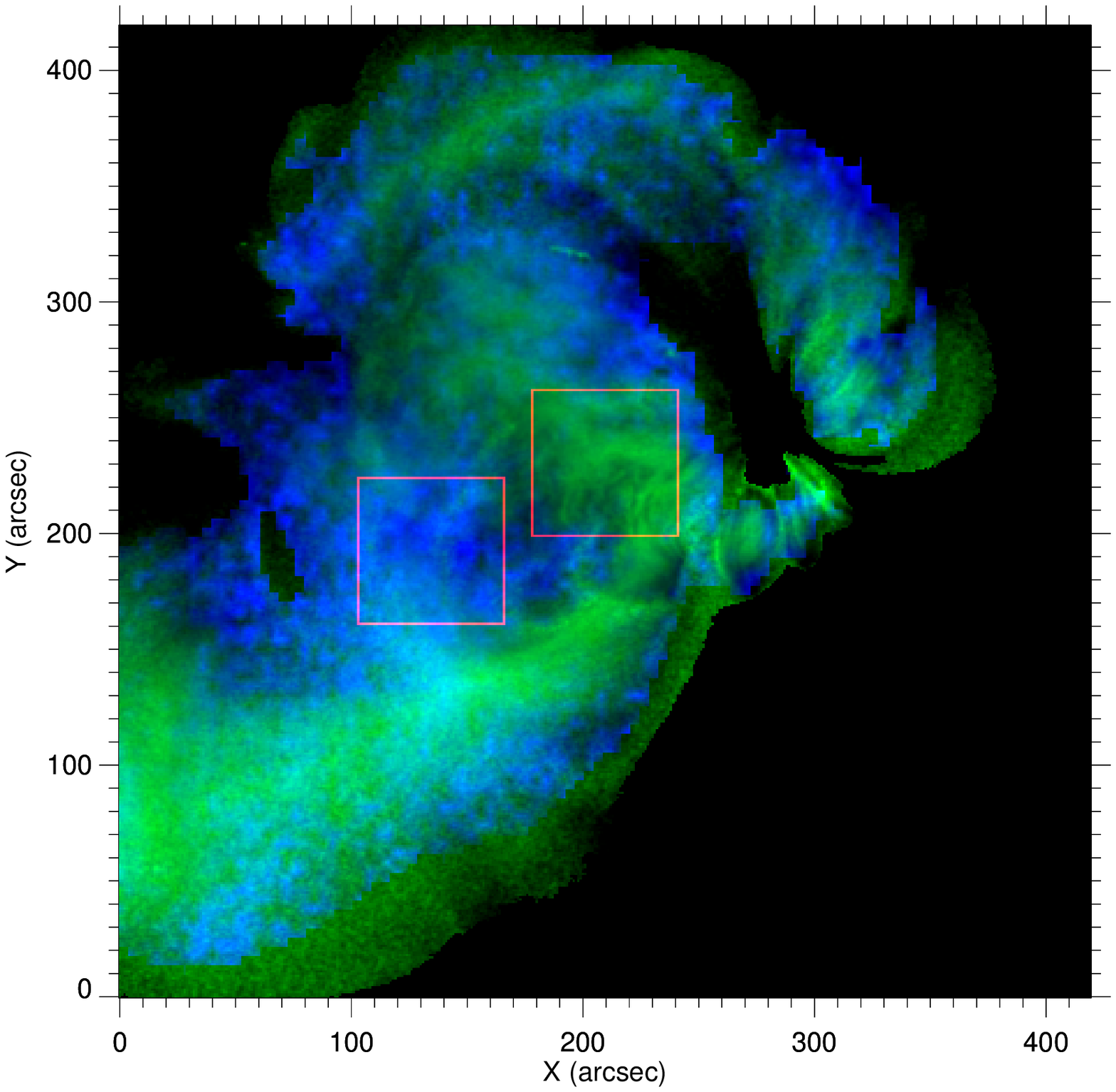}
  \caption{Maps of temperature as in Fig.~\ref{fig1} obtained with CIFR
  (green scale) and hard (F4/F5) filter ratio (blue scale). The red boxes
  mark the regions of different temperature regimes analyzed separately:
  the one on the left is hotter in the hard filter ratio (hereafter hard-hot region ), and the
  one on the right is hotter in the soft filter ratios (hereafter soft-hot region).
 }
  \label{fig:mapsub}
\end{figure*}

Important information comes from the so-called Emission Measure vs
Temperature diagram (EM(T), e.g. \cite{2000ApJ...528..524O}). This diagram
is a histogram of the distribution of the emission measure in
temperature bins. To build this histogram we consider the maps of
temperature and emission measure obtained with a given filter ratio
and sum the emission measure of all pixels with a temperature in a
certain bin. Fig.~\ref{fig:emt_ar} shows EM(T) obtained from the
relevant filter ratios, including CIFR; the bin size is the same as
the temperature resolution of the filter response functions, i.e.
$\Delta \log T = 0.1$. The distributions obtained with soft simple
filter ratios and with CIFR mostly overlap: they are strongly peaked
at $\log T \approx 6.4$ and fall by about three decades in a
temperature range $6.2 < \log T < 6.7$. EM(T) obtained with both the
intermediate filter ratios F4/F3 and F5/F3 are similar to those from
the soft filters, but differ in several respects:  they are slightly
shifted to higher temperatures (peaking at $\log T \approx 6.5$);
they have a somewhat smaller amplitude; and they decrease with
temperature less slowly on the hot side (decreasing by 2.5 decades
at $\log T = 6.8$). EM(T) obtained from the hard filter ratio F4/F5
peaks at the same temperature and with the same amplitude, but its
hot tail extends to $\log T > 7$ still at about a few percent of the
peak. The error in each EM value is negligible in this scale.

Our question is now whether the hot component found with the hard
filter ratio is real or just, for instance, an artifact of higher
uncertainties due to the shallow dependence of this ratio on
temperature and to the lower photon statistics.

In order to investigate this issue, we first recall from
Fig.~\ref{fig1} that regions of enhanced temperature occur at
different locations in the soft and hard filter ratio maps. The maps
appear anti-correlated. Now we identify two 64x64 pixel (16x16 boxes in F4/F5 maps) subregions
that are reasonably homogeneous, as indicated in
Fig.~\ref{fig:mapsub}. The one to the left has enhanced temperatures
in the hard ratio map, and the one to the right has enhanced
temperatures in the soft ratio maps.  We refer to these as the
hard-hot (HH) and soft-hot (SH) regions, respectively. 

\begin{figure}[!t]
  \centering \includegraphics[width=8cm]{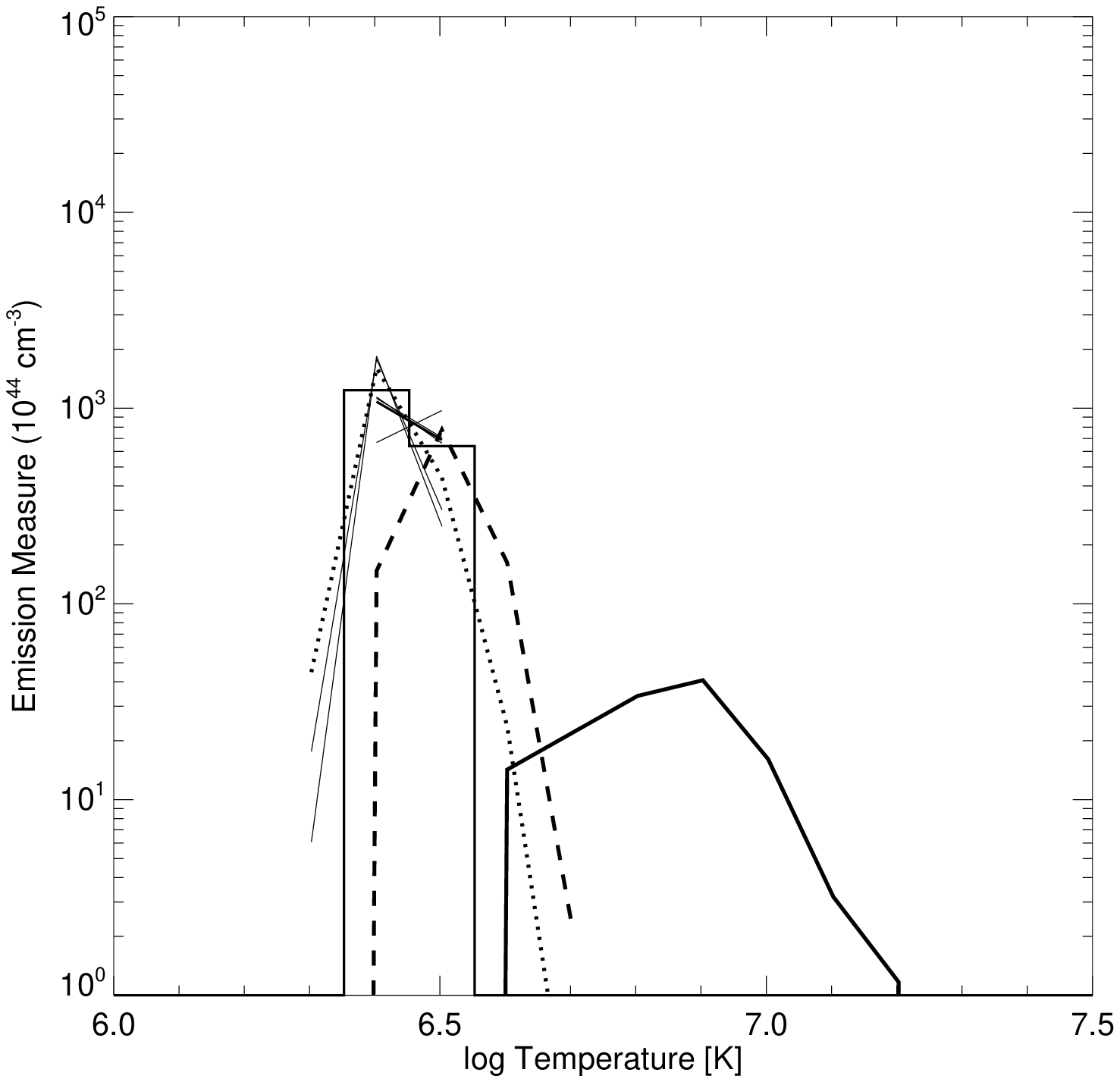}  \includegraphics[width=8cm]{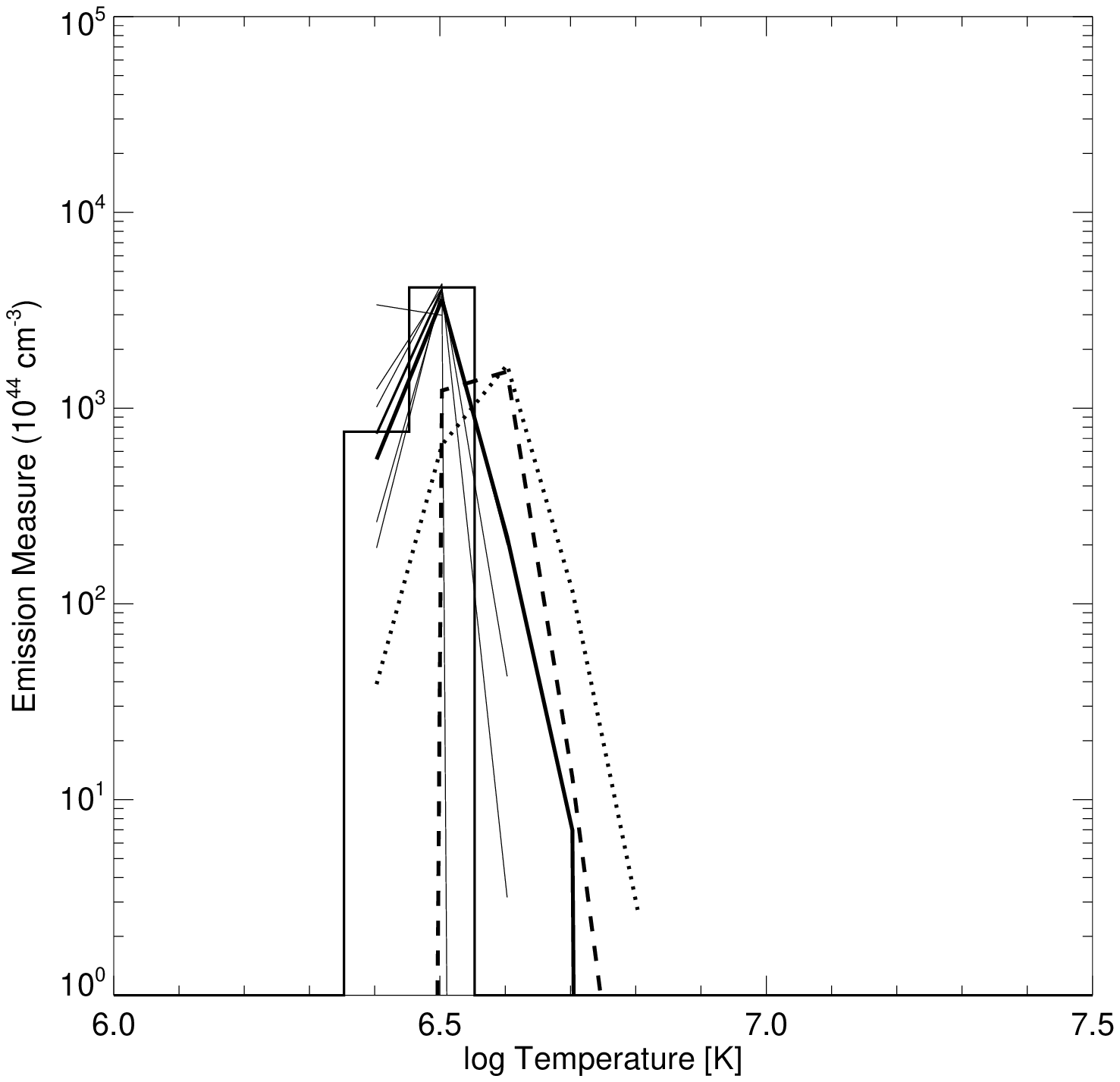}
  \caption{Emission measure distributions vs temperature for the hard-hot (left) and soft-hot (right) subregions marked in Fig.~\ref{fig:mapsub}. Lines as in Fig.~\ref{fig:emt_ar}.
 }
  \label{fig:emt_reg}
\end{figure}

The EM(T) distributions obtained separately for the two subregions
are shown in Fig.~\ref{fig:emt_reg}.  We immediately see that they
are very different from each other.  They are also qualitatively
different from the distribution for the whole active region. Their
amplitude is much smaller because they include much less plasma. 

EM(T) of the soft-hot region (right panel) is narrow in {\it all} filter
ratios, including the hard ratio. None of the distributions has a
hot tail, with all dropping to essentially zero by $\log T =6.8$.
The peaks of the distributions are all clearly at a higher
temperature ($\log T \approx 6.5 - 6.6$) than those of the whole
active region ($\log T \approx 6.4 - 6.5$).

EM(T) of the hard-hot region (left panel) is different in several
aspects. First, EM(T) of the hard ratio is considerably hotter than
all the others: it is quite broad and its peak is at $\log T \approx
6.9$ with a hot tail extending beyond $\log T = 7$.  It appears to
be detached from the other distributions, and its amplitude is
smaller by a factor of 50. The EM(T) of the other filter ratios are
consistently narrow and peaked at $\log T \approx 6.4 - 6.5$.  This
is slightly cooler than the peaks in the soft-hot region.  The small
but clear temperature difference is highly significant, as we will
show. It explains the anti-correlation that is evident in the hard
and soft temperature maps of Figure \ref{fig1}.  The key result is that
in regions where the hard ratio extends to high temperatures (e.g. left frame in Fig.~\ref{fig:mapsub}) the soft ratios peak at slightly lower temperatures than in regions where the hard ratio is confined to cooler temperatures (right frame in Fig.~\ref{fig:mapsub}).

It is important to remember that the plasma is multi-thermal along
each line-of-sight, whereas our EM(T) distributions were obtained by
assuming that it is isothermal.  Important questions are therefore:
can we combine the information from our measurements into a coherent
picture? Are filter ratio
diagnostics reliable? Do they provide sensible information? Is hot
plasma really hot or are we being fooled by large errors in T
associated with the shallow ratio vs T curve of the F4/F5?

\subsection{MonteCarlo Simulations}

There are multiple nonlinear effects that complicate the derivation
of temperature maps and emission measure distributions.  These
include the nonuniform thermal distribution along the line of sight,
its variation from pixel to pixel, the differing sensitivities of
the filters, and the non-linear weighting of the thermal components
through the filter responses.  In order to assess quantitatively all
such effects we perform MonteCarlo simulations including all
realistic ingredients.

Our approach is to create synthetic images of approximately
homogeneous sub-regions similar to the hard-hot and soft-hot regions
of Figures \ref{fig:mapsub} and \ref{fig:emt_reg}, and then to generate 
EM(T) distributions exactly
as we did for the real data.  We take the sub-regions to be
64$\times$64 pixels in size, and we assume that EM(T) along the
different lines of sight within each sub-region are variations on
the same ``parent" distribution EM$_{los}$(T).  The parent distribution has a
parameterized form.  We vary one or more of the parameters at each
pixel by randomly sampling from a probability distribution (either
normal or log-normal). This gives us an input EM(T) at each pixel
from which we compute intensities corresponding to observations made
with the different filters.  As a final step, we modify the
intensities with Poisson noise to mimic photon counting statistics.

The EM$_{los}$(T) parent distributions are either single or double step
functions (top-hat functions).  The amplitude ($\sigma_{EM}$) and
central temperature ($\sigma_{[\log T]}$) of the step function can vary,
but its width ($d \log T$) is fixed for the entire sub-region
simulation. 

From the parent EM$_{los}$(T) we compute the corresponding emission value for each of the five filters. We then randomize the emission values, according to Poisson statistics on photon counts. The DN rate is first integrated on the exposure time to obtain the total DN counts. The DN counts are then converted to photon counts, which are randomized and then converted back to DN counts and DN rate.
The 5 new DN rates are assigned to a pixel and the whole procedure is repeated 64$\times$64 times to build a 64$\times$64 pixel image.

The resulting image is analyzed as done for the real XRT images, i.e. derive temperature and emission measure maps with the filter ratio method and build the related EM(T). We apply the same rebinning, the same constraints and the same acceptance thresholds as those used for the real data. We obtain an EM(T) distribution for each filter ratio.

\begin{figure*}[!t]
  \centering \includegraphics[width=8cm]{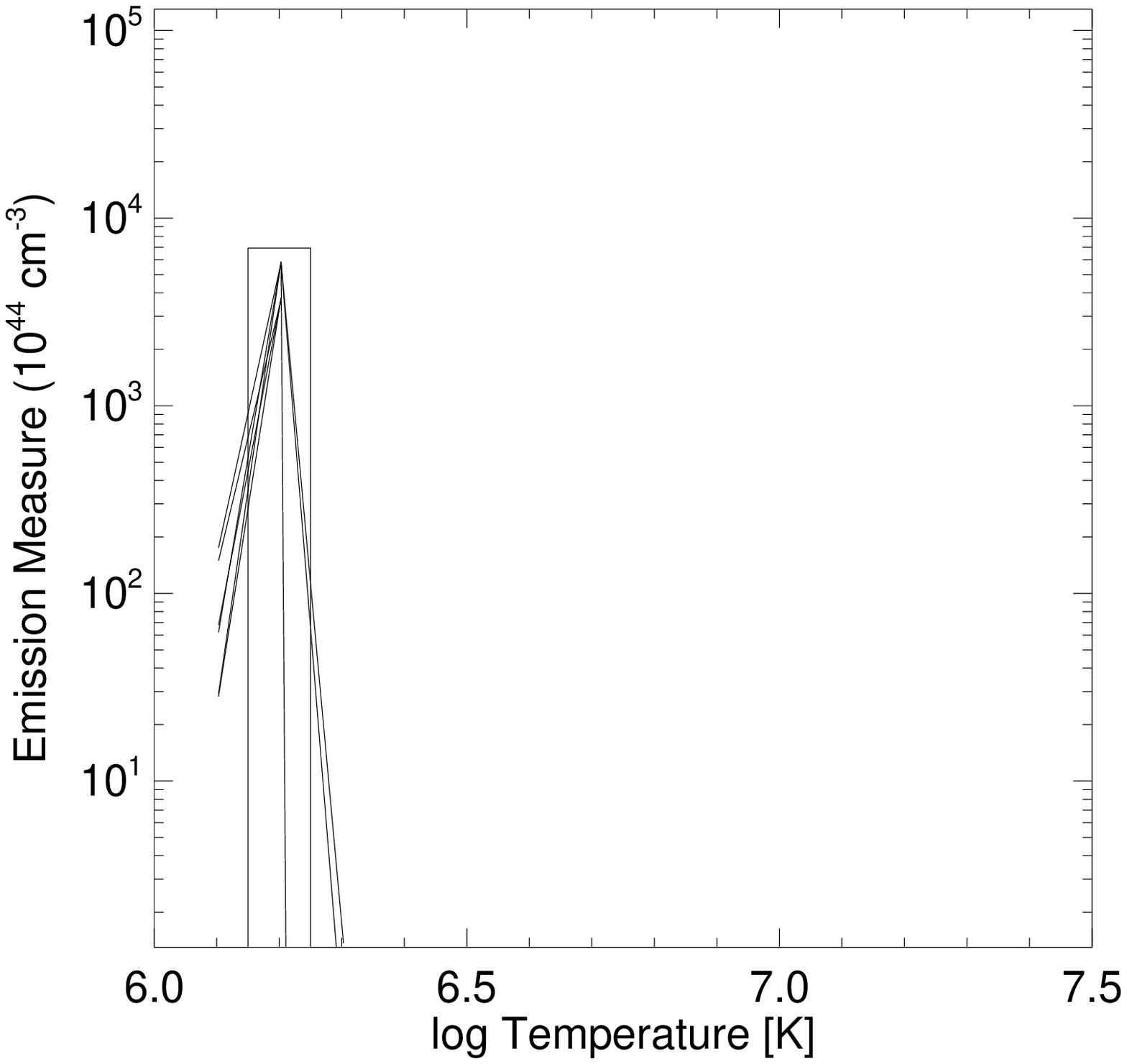} \includegraphics[width=8cm]{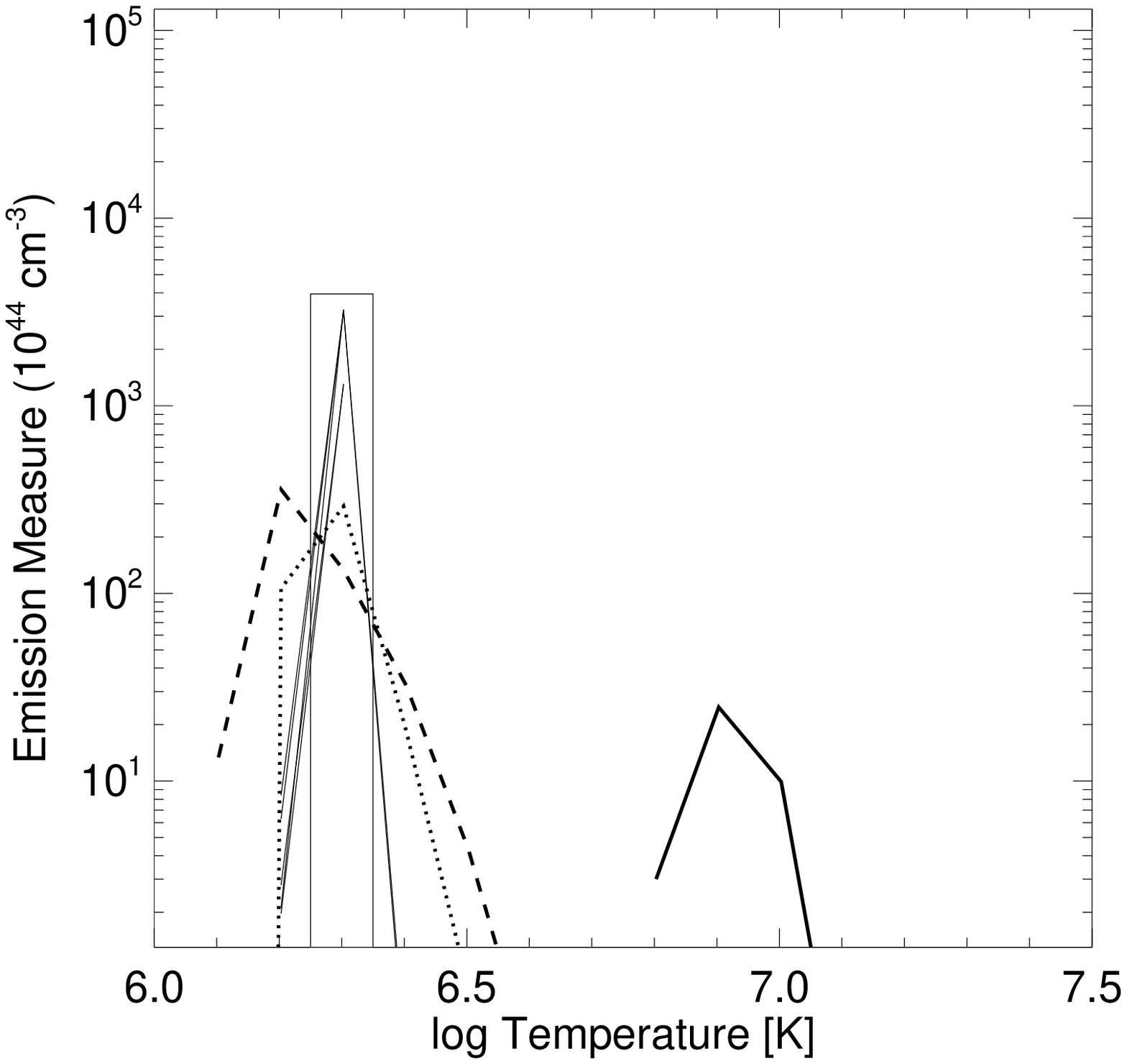}
  \centering \includegraphics[width=8cm]{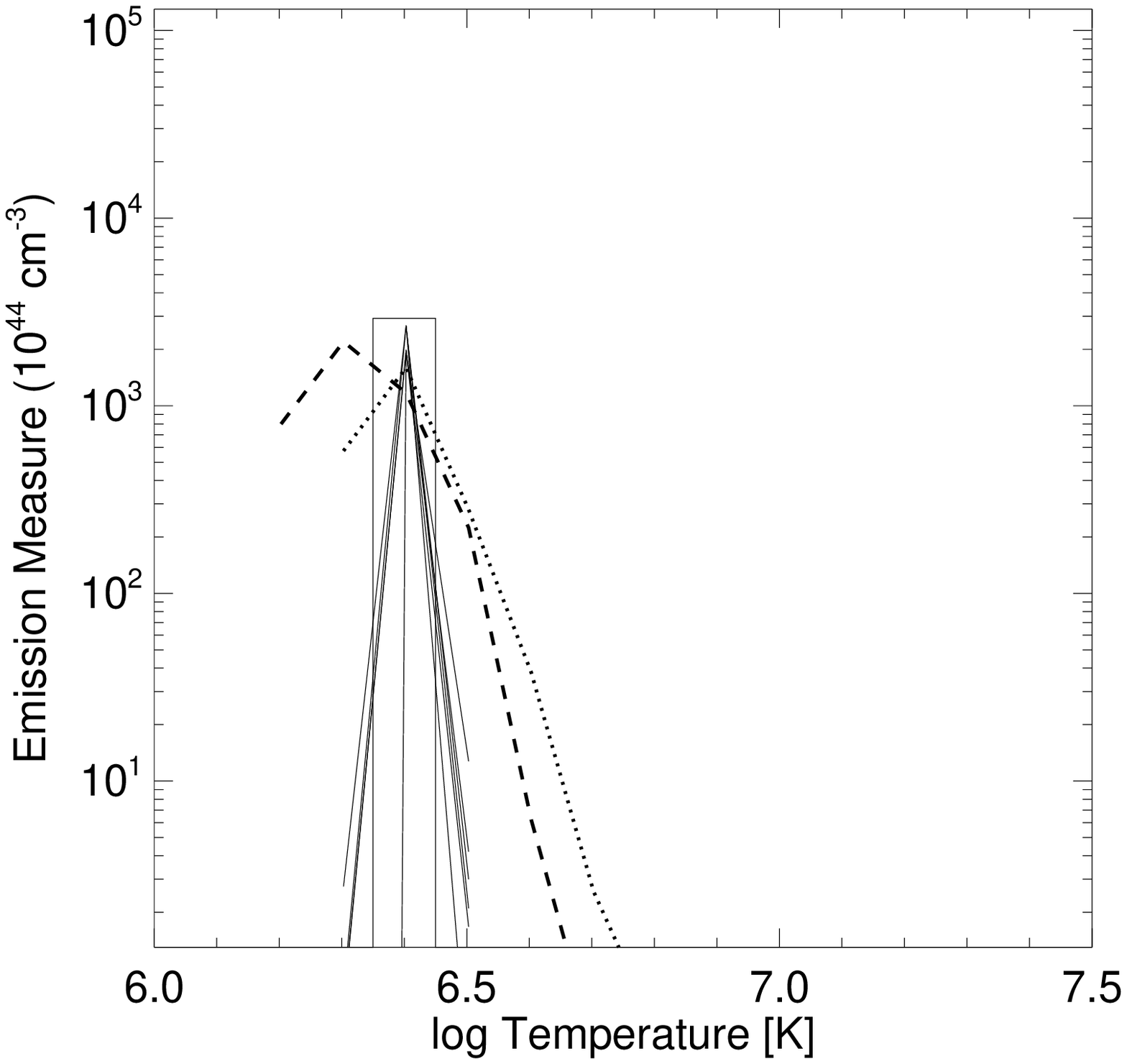} \includegraphics[width=8cm]{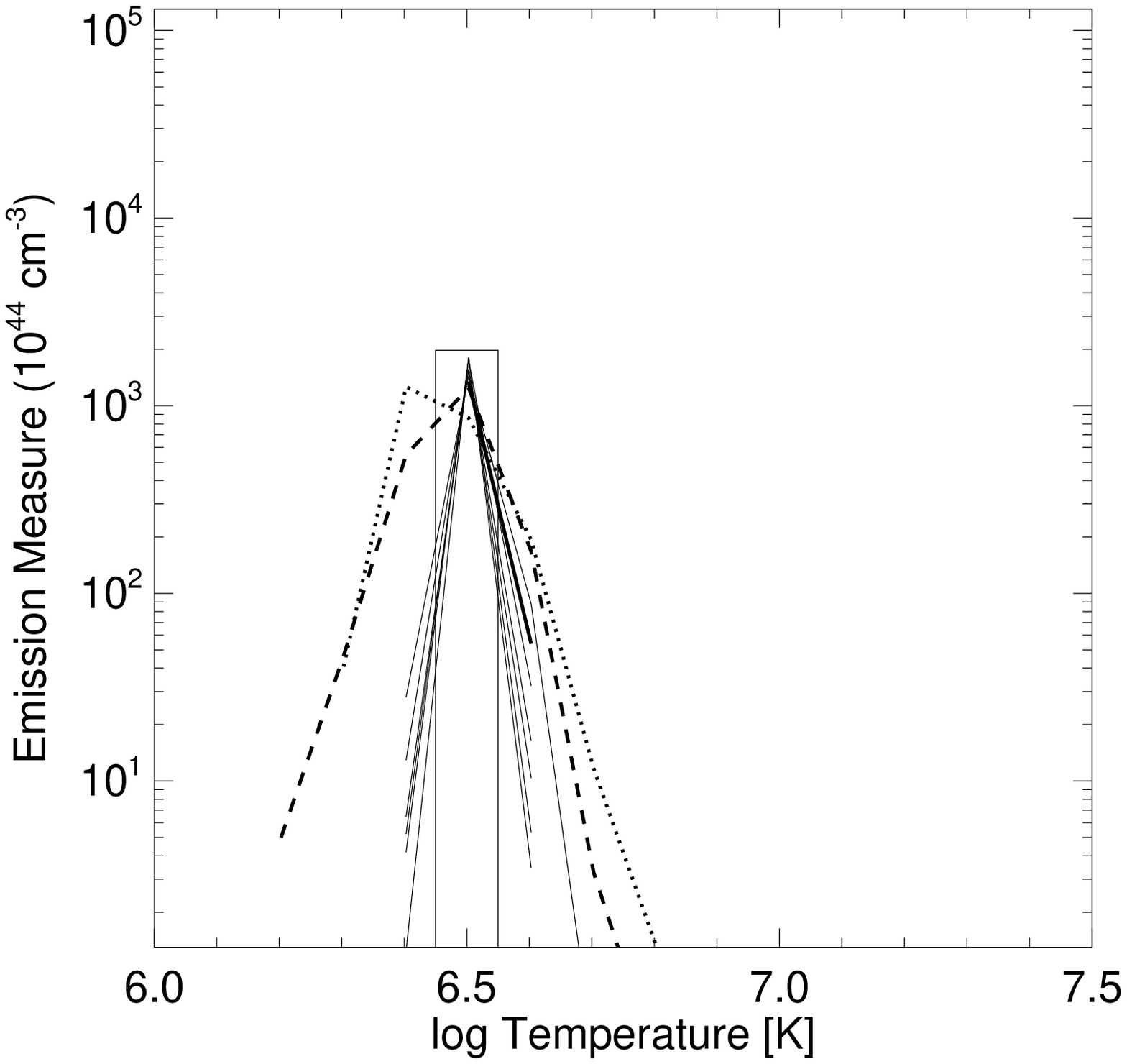}
  \caption{Emission measure distributions vs temperature obtained from MonteCarlo simulations of images in the five XRT filters of the active region observation. We show the parent EM(T) ({\it histogram}), and those obtained in the same filter ratios as those in Fig.~\ref{fig:emt_ar} and ~\ref{fig:emt_reg} (same lines), except for CIFR. From left to right and top to bottom it is assumed a single component parent EM(T) centered at $\log T = 6.2, 6.3, 6.4$ and 6.5, respectively and a width $d \log T = 0.1$. The EM(T) is randomized only for the amplitude with $\sigma = 30$\%.
 }
  \label{fig:em1}
\end{figure*}

\begin{figure*}[!t]
  \centering \includegraphics[width=8cm]{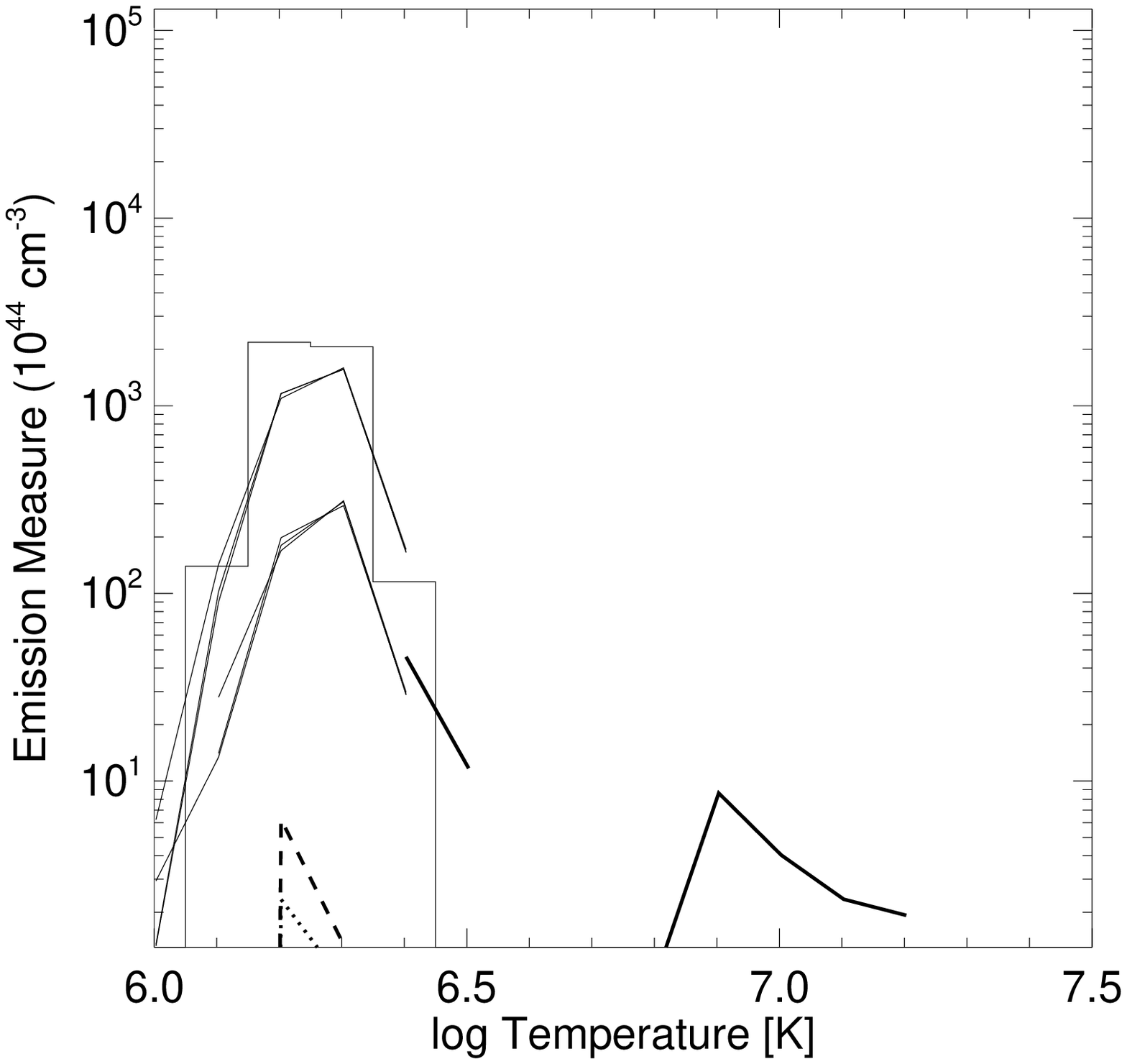} \includegraphics[width=8cm]{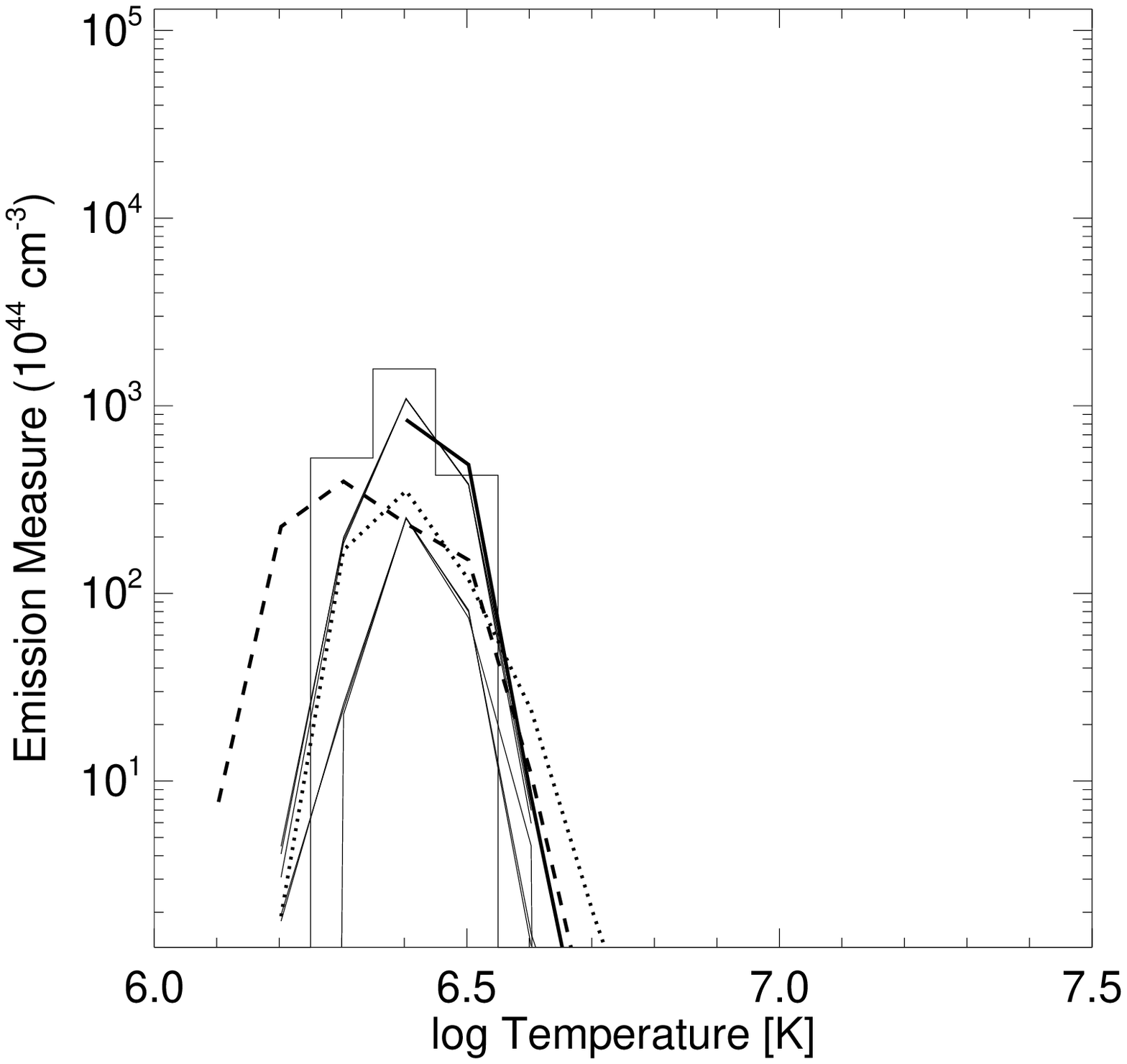}
  \caption{As Fig.~\ref{fig:em1} but randomizing the EM(T) also for central temperature with $\sigma_{[\log T]} = 0.1$ around $\log T = 6.2$ and with $\sigma_{[\log T]} = 0.05$ around $\log T = 6.4$.
 }
  \label{fig:em2}
\end{figure*}

\begin{figure*}[!t]
  \centering \includegraphics[width=8cm]{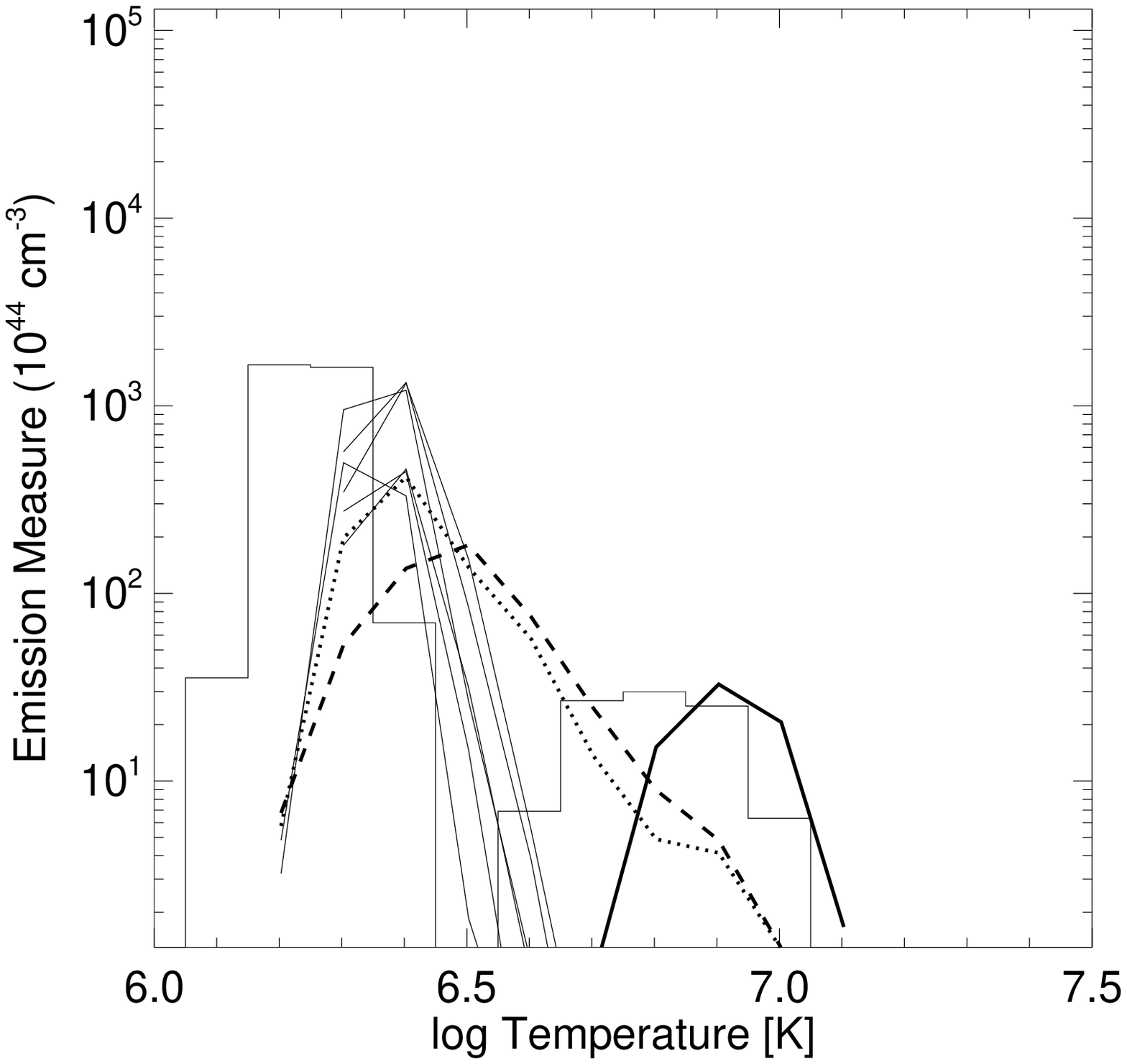} \includegraphics[width=8cm]{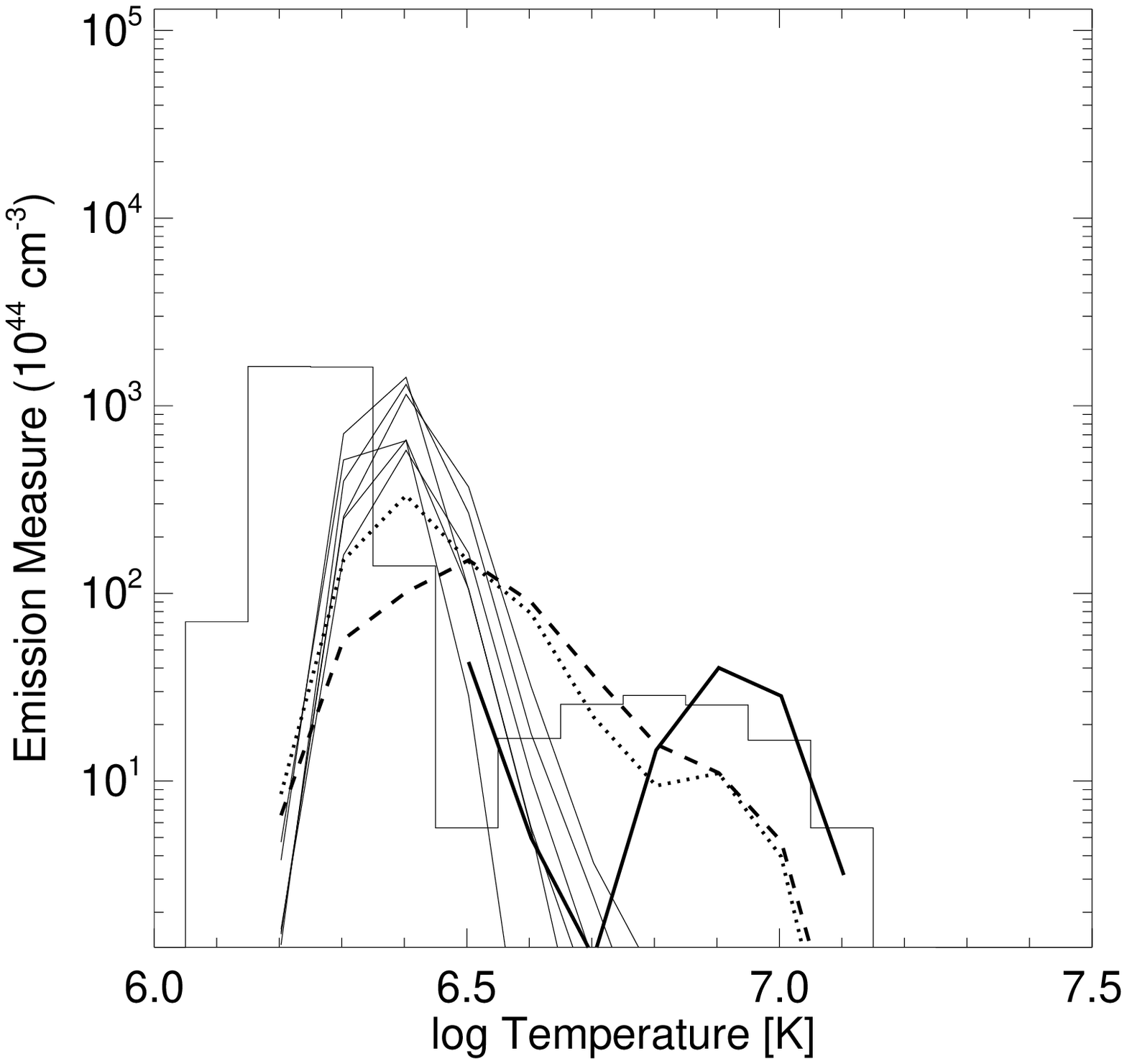}
  \caption{As Fig.~\ref{fig:em2} but including a second hot component in the parent EM(T) centered at log T = 6.8 randomized with  $\sigma_{[\log T]} = 0.05$, and with width $d \log T = 0.15$ (left) and with  $\sigma_{[\log T]} = 0.1$, and with width $d \log T = 0.20$ (right).
 }
  \label{fig:em3}
\end{figure*}

\begin{figure*}[!t]
  \centering \includegraphics[width=8cm]{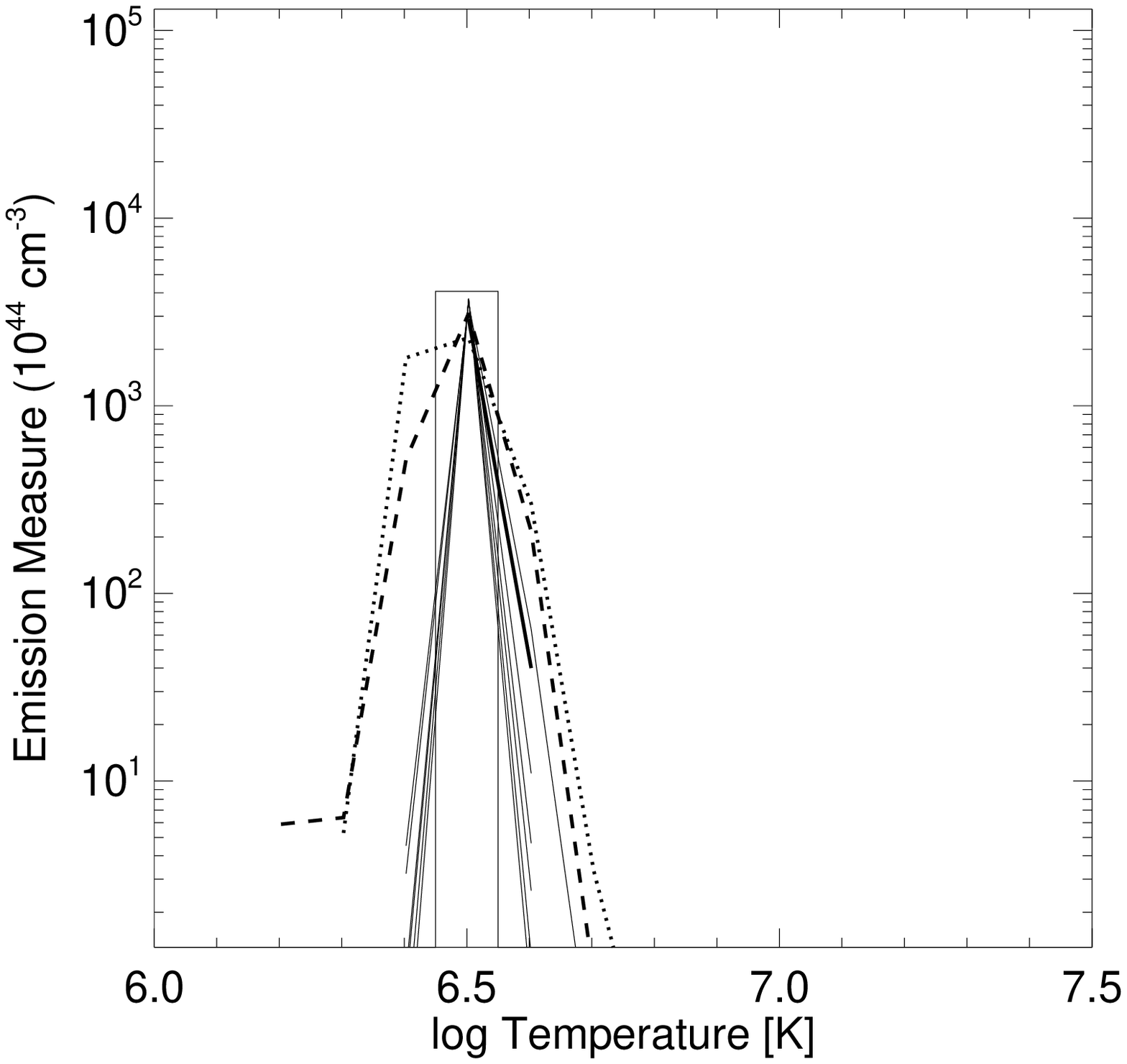} \includegraphics[width=8cm]{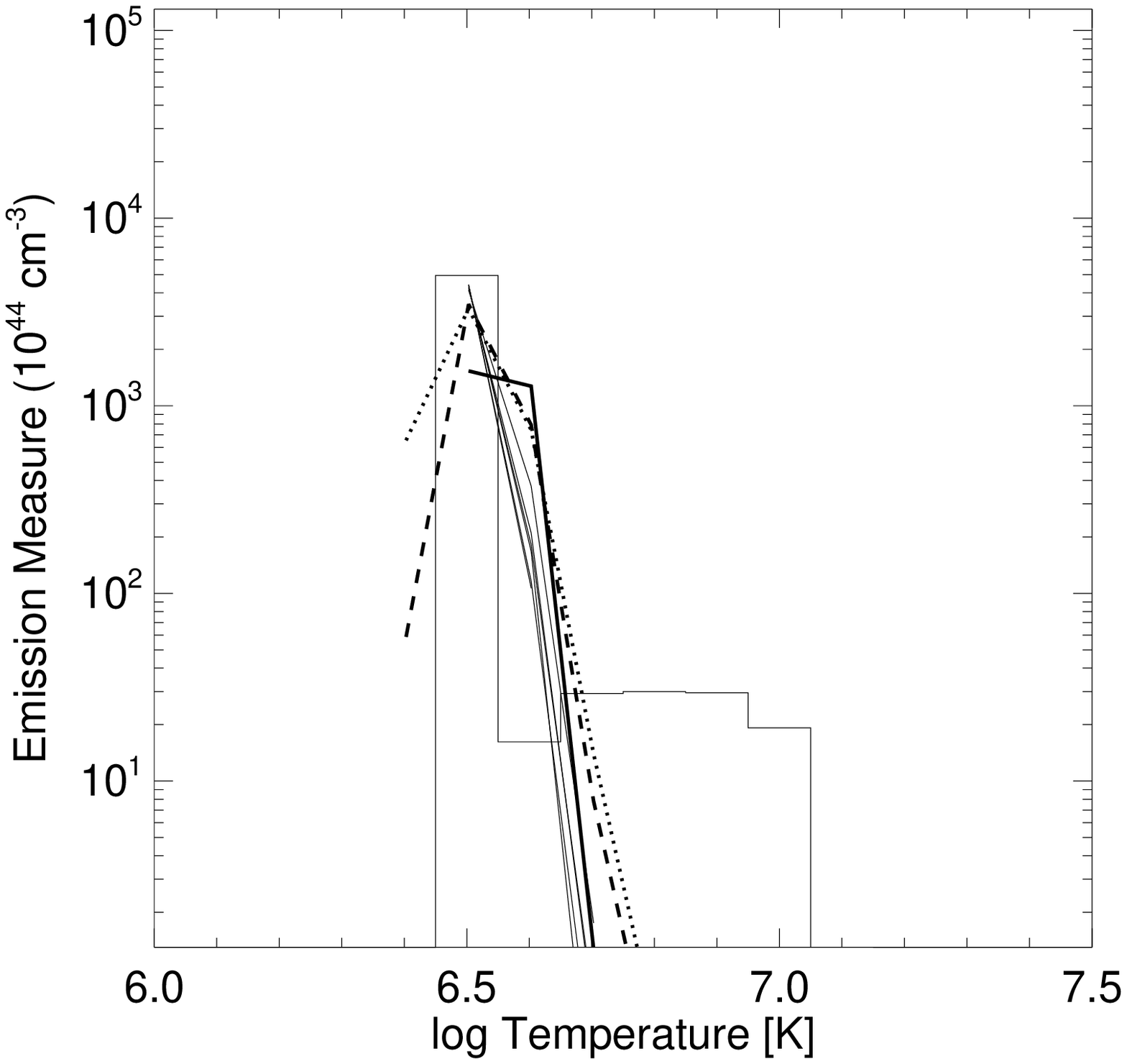}
  \caption{As Fig.~\ref{fig:em1} but showing comparison between the EM(T) obtained from a parent single-component centered at log T = 6.5 (left) and the same parent EM(T) with the addition of a smaller hot component as in Fig.~\ref{fig:em3}(right).
 }
  \label{fig:em4}
\end{figure*}

We show end results of this procedure in
Figs.~\ref{fig:em1},\ref{fig:em2},\ref{fig:em3},\ref{fig:em4} in a
format similar to -- and with the same resolution as -- that used in
Figs.~\ref{fig:emt_ar},\ref{fig:emt_reg}. The histogram in each
panel is the total input EM$_{los}$(T) for the sub-region, i.e., the
modified (randomized) EM$_{los}$(T)'s for all the pixels summed together.
The connected points are the EM(T) obtained from the simulated
observations of this input.  For the sake of clarity we do not show
the EM(T) obtained with CIFR, which is invariably very similar to
those obtained with the soft filter ratios (the histogram is now the
total input EM$_{los}$(T)).

Fig.~\ref{fig:em1} shows results for single component parent EM$_{los}$(T)'s
with four different central temperatures, increasing from $\log T =
6.2$ in the top-left to $\log T = 6.5$ in the bottom-right.  Only
the emission measure amplitude and photon counting errors vary
within each case. The amplitude is normally distributed with a
Gaussian half width $\sigma_{EM} = 30$\%. Therefore the total input EM(T) has the same shape as the parent EM$_{los}$(T) in each pixel. The DN rate
level is set so as to be comparable to the one observed in the soft
filters (F1 and F2) in the left region in Fig.~\ref{fig:mapsub}.

For $\log T = 6.2$ (upper-left) only the soft filter
ratios yield meaningful temperature and EM values, which closely
resemble the input EM. There is a small spread due only to photon
and EM amplitude fluctuations, and not to temperature variations. For $\log T = 6.3$ (upper-right), the
soft filter ratios and one of the medium ones still reproduce the
input EM$_{los}$(T) quite well. The other medium filter ratio yields a
somewhat broader distribution peaked at lower temperature. Due to
random photon counting errors, the hard filter ratio produces a
spurious faint hot component centered at $\log T = 6.9$. Although
this hot component has some resemblance to that observed in the left
panel of Fig.~\ref{fig:emt_ar}, the agreement is in
fact not good because the cool components are cooler than the
observed ones. For $\log T = 6.4$ and $\log T = 6.5$ we once again
obtain single EM(T) components in all filter ratios, mostly centered
at the center of the parent EM$_{los}$(T). 
The spurious hot component in the hard ratio disappears because of
the improved photon counting statistics afforded by the hotter input
distribution.

Fig.~\ref{fig:em2} shows the effect of randomizing the central EM(T)
temperature.  This is done for two single-component parent EM$_{los}$(T)'s
at different central temperatures:  $\log T$ = 6.2 and 6.4 in the
left and right panels, respectively. Pixel-to-pixel variations in
the central temperature are taken from log-normal distributions with
a half-widths $\sigma_{[\log T]} = 0.1$ and 0.05,
respectively. The effect is clearly to broaden the soft EM(T)'s.  A
spurious hot component appears in the hard filter ratio, but only
for the cooler case, again due to poor photon counting statistics.
An important conclusion is that observed distributions like those in
the left panel of Fig.~\ref{fig:emt_reg} cannot be produced by single component
input distributions: high temperature values in the hard ratio can only
occur when the emitting plasma is all cooler than diagnosed in the soft ratios.

In Fig.~\ref{fig:em3} two examples show the effect of adding the
second hot component to the input EM$_{los}$(T). The stronger cool component
has the same parent EM$_{los}$(T) as shown on the left of
Fig.~\ref{fig:em2}. The parent EM$_{los}$(T) of the hot component, another
step function, is centered at $\log T= 6.8$ and has a width of
$d\log T = 0.15$ on the left and 0.2 on the right. The input
distribution is broadened further by randomizing the central
temperature with $\sigma_{[\log T]} = 0.05$ and 0.1, respectively.  The
figure shows that the presence of the second hot and weaker
component naturally leads both to an increase of the temperature of
the EM(T) from the soft filter ratios and to the detection of a
single peaked hot component from the hard filter ratio. The medium
filter ratios F5/F3 and F4/F3 both produce a distribution with a
peak around $\log T = 6.4 - 6.5$ but with a hot tail extending to
$\log T \approx 7$. The distributions in the soft and hard filter
ratios look very similar to the ones observed from the hard-hot
subregion on the left in Fig.~\ref{fig:emt_ar}.  The central temperatures,
amplitudes, and widths all agree well. The hot tails predicted for
the medium filter ratios is not observed, however.

In our opinion Fig.~\ref{fig:em3} provides the best approximations
to the EM(T) derived from the observed hard-hot subregion, at least
for the types of input distributions that we have so far explored.
In particular, they produce a hot component in the hard ratio at the
same time that they produce a cool component with sufficiently high
temperature (peaking at $\log T \approx 6.4$) in the soft ratios.
These are important constraints that must both be satisfied.  The
total emission measure summed over temperature also
agrees well with the observations, i.e., to within 20\% for all
filters except for F3 ($\sim 30$\%).

We now turn our attention to the soft-hot subregion, whose EM(T) is
shown on the right of Fig.~\ref{fig:emt_reg}. In contrast with the
hard-hot subregion, all of the filter ratios, including the hard
ratio, give a single quite narrow EM(T) distribution peaking at
$\log T \approx 6.5$. The left panel of Fig.~\ref{fig:em4} shows that
this distribution can be recovered from a narrow input EM$_{los}$(T) with
the same peak temperature. However, the right panel shows that it
can also be recovered if the input EM$_{los}$(T) contains a weaker hot
component. The hot component is not detected even by the hard ratio
because the much stronger cool component is hot enough to dominate
the observed signal. This would not be the case if the cool
component were slightly cooler, because the sensitivity of the hard
filters is a steep function of temperature in this range.  The hard
filters are nearly blind to plasma cooler than $\log T = 6.4$, and if
this plasma were invisible, the hard ratio would be dominated by the
hot component, even though it is inherently weaker (Fig. \ref{fig_rsp}). 

Thus, the data are consistent with a picture in which a weak hot
component is present everywhere, and the only significant difference
between hard-hot and soft-hot type subregions is that the cool
component is slightly cooler in the hard-hot subregions.

\section{Discussion}

In this work we have analyzed an active region observed by Hinode/XRT in November 2006, i.e. in the first phase of the Hinode mission. The observation is made with five filters which can be combined into ten different but not entirely independent filter ratios to provide temperature diagnostics. We have optimized the signal-to-noise ratio by averaging over one hour of observation, i.e. 12 images per filter, in the absence of flaring activity and any other considerable variation. 
The new achievement is to include in the analysis  the ratio of the filters of intermediate thickness (Be\_med, F4, and Al\_med, F5) -- the thickest ones available in this observation -- which screen out the emission of plasma below $\log T = 6.3$ and are sensitive therefore to possible weak hotter plasma components. We have derived temperature and emission measure maps for all filter ratios (except for the softest one) for signal above a high acceptance level. By combining the values obtained at individual pixels, we have built emission measure distributions as a function of temperature which show evidence of a hot plasma tail up to $\log T > 7$. From inspection of the temperature maps we realize that there is a clear correlation between regions detected as hot in the soft filter ratios and regions detected as cool in the hard filter ratio and viceversa (i.e. an anticorrelation). To investigate this point further we have first extracted the emission measure distribution from two regions of the different kind. The distribution of the former type is characterized by a single narrow component with the peak at $\log T \approx 6.5$ in all filter ratios. The distribution of the latter type is instead split into a cool component peaking at $\log T \approx 6.4$ detected by the soft filter ratios and a hot component at $\log T \approx 6.9$ detected by the hard filter ratio. In order to interpret this result correctly we have performed MonteCarlo simulations to build realistic synthetic XRT images in all 5 filters starting from model parent emission measure distributions along the line of sight. We find that both kinds of emission measure distributions derived from the data can be coherently explained with a single type of line-of-sight parent distribution consisting of a high cooler component at $\log T \leq 6.5$ and lower hotter component extending to $\log T \sim 7$. The difference between obtaining a single or split distribution is determined by the peak temperature of the cool component. If it is below $\log T = 6.3$ the hard filter ratio is able to detect the hot component, otherwise the cool component invariably obscures the other one.

We are unable to obtain a perfect match between simulations and data results for a variety of reasons. First, although we have performed a fine tuning of the hard filters calibration through detailed feedback on the data there is still room for some uncertainty. In particular, it is possible that the thickness of one of the intermediate filters is even smaller than obtained from the tuning, i.e. we have applied the minimum possible correction to obtain consistent results. This assumption is conservative and we may obtain even hotter emission measure component after applying possible less conservative corrections. Small calibration tuning on the F3 (Be\_thin) filter response may improve the agreement of results from ratios involving this filter. Second, we have assumed very simplified forms of the parent emission measure distribution along the line of sight. The results are clear as such and do not require more refined forms. In any case we prefer here not to push the temperature resolution of our emission measure analysis at any step beyond the one offered by the filter calibration, i.e. 0.1 in $\log T$. This includes also the randomization of the parameters. Finally, our analysis does not account for the inherent organization of the plasma in coherent magnetic loops. We do not pretend here to enter into the fine details of the emission measure distribution.

\begin{figure*}[!t]
  \centering \includegraphics[width=12cm]{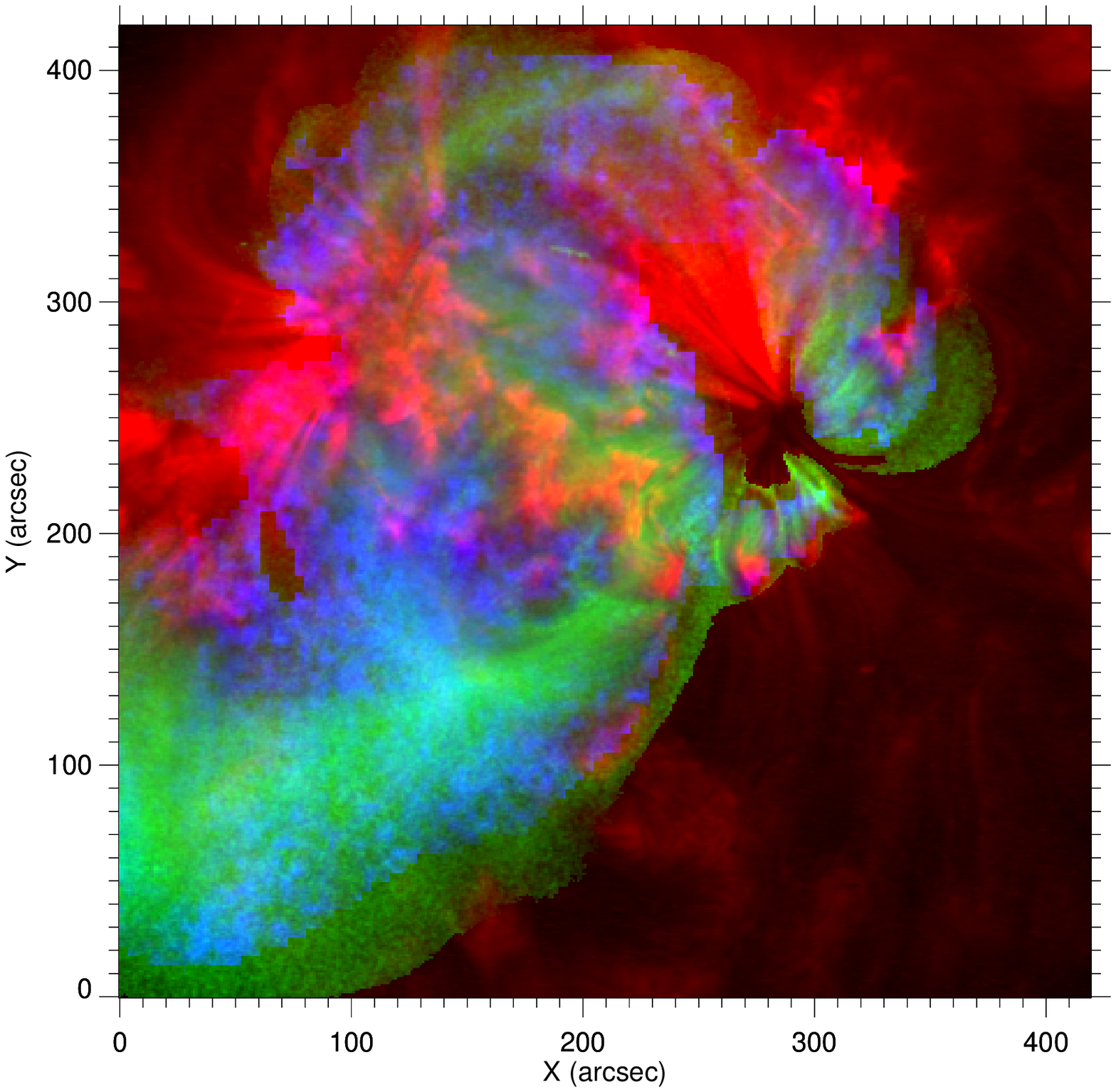}
  \caption{True color image showing the temperature map obtained with CIFR (green scale), with the hard (F4/F5) filter ratio (blue scale) and the emission map detected with the TRACE 171 \AA filter band (red scale).
 }
  \label{fig:maptrue}
\end{figure*}

Going back to the morphology of the active region, we propose that hot plasma is widespread in the active region but that its detection depends somehow inversely on the local level of activity. In particular, the hot plasma may be difficult in high pressure loops, where the dominant cool component is relatively hot, and easier to detect in fainter structures, where the dominant component is relatively cool. This scenario is consistent with what we find when we overlay the temperature maps obtained with the soft and hard filter ratios with an emission map obtain with the 171 \AA filter onboard TRACE, which is sensitive to plasma at $\sim 1$ MK (Fig.~\ref{fig:maptrue}). We immediately realize that the central green loops (in the soft-hot subregion) are anchored to moss structures detected by TRACE which are typically interpreted as the footpoints of high pressure loops (e.g.
\cite{1994ApJ...422..412P,1999ApJ...520L.135F}). Longer loops detected by TRACE are clearly cool structures which occupy volumes complementary to the hotter structures detected with XRT. Hot (blue) regions are generally faint localized at the boundaries.

Our analysis has therefore revealed clear evidence for small quantities of very hot plasma that is widely distributed throughout the active region. The diagnosed temperatures are characteristic of flares ($\sim 10$ MK) though no flares occurred within about 2 hrs of the observations. This provides strong evidence for the nanoflare picture of coronal heating.  The emission measure of the hot components ($\log T \geq 6.6$) is about 3\% of the cool one in both parent EM$_{los}$(T)'s of Figs.~\ref{fig:em3} (right) and \ref{fig:em4} (right). We mention that similar results are obtained from the analysis of the same active region observed 2 days before (10 November), in a period when the region was even more quiet.
We have run nanoflare simulations with our Enthalpy-Based Thermal Evolution of Loops (EBTEL) code \citep{2008ApJ...682.1351K} and have no difficulty producing emission measure distributions with this property. 
A similar EM distribution has been also obtained in \cite{2008ApJ...684..715R} with a simulation of a loop heated by nanoflares localized at the footpoints and lasting 3 min, taking deviations from ionization equilibrium into account.

The standard nanoflare model envisions that many unresolved strands at different stages of heating and  cooling should coexist within close proximity of each other.  It therefore predicts a broad emission measure distribution along each line of sight.  (The distribution could be rather narrow for a bright coronal loop that is heated by a storm of nanoflares confined to a short time window).  Should nanoflares be ruled out in those parts of the active region where we find no detectable signal in the hard ratio?  They cannot be ruled out for the following reason.

The hard filters are much more sensitive to hot plasma than to cool plasma.  For example, the sensitivity of F5 is 10 times greater at 10 MK than at 3 MK.  As noted above, however, nanoflare simulations predict an emission measure that is at least 100 times smaller at the hotter temperature.  The observed signal will therefore be dominated by the cooler plasma.
Even with the hardest filter, there is a tendency for small but significant amounts of hot plasma to be masked by more dominant cool plasma.

The hot component is a minor component of the total emission measure distribution, much smaller (about 3\%) than the cooler $\sim 3$ MK dominant component. The small emission measure of the hot component explains why it has been so elusive so far: it is overwhelmed by the cooler component along the line of sight and can be detected only after cutting the cool component off. This screening is efficiently but not completely performed through the single thick filters: the EM(T) distribution obtained with the ratio of the filters shows that most but not all of the dominant 3 MK plasma is masked out. The final cut is allowed by the temperature diagnostics provided by the filter ratio.

In conclusion, the thick filters of Hinode/XRT detect hot flare-like plasma in an active region outside of proper flares. This plasma is widespread and steady, and its amount is consistent with that predicted by nanoflare models, which receive great support here as candidate to explain coronal heating.

\bigskip
\acknowledgements{We thank G. Peres and P. Grigis for useful suggestions. Hinode is a Japanese mission developed and launched
by ISAS/JAXA, with NAOJ as domestic partner and NASA and STFC (UK) as
international partners. It is operated by these agencies in co-operation
with ESA and NSC (Norway).  FR acknowledges support from
Italian Ministero dell'Universit\`a e Ricerca and Agenzia Spaziale Italiana
(ASI), contract I/015/07/0. PT is supported by
NASA contract NNM07AA02C to SAO. JAK was supported by the NASA Living With a Star program. SP acknowledges the support from the Belgian Federal Science Policy Office through the ESA-PRODEX programme. This work was partially supported by the International Space Science Institute in the framework of an international working team.}

\appendix

\section{Fine tuning of filter calibration}

In this paper we use the recently updated XRT filter thickness calibration \citep{Naru09}.  The high quality of the XRT observations here
analyzed allows to test this new instrument calibration. While changes
in transmissivity with respect to the old calibration are in general
limited for all filters used here (of the order of up to about 15\%), the effect on
the temperature and emission measure diagnostics can be significant. This
is particularly true in the case of the ratio of hard filters (F4/F5)
which spans a rather narrow range compared to other filter ratios.

\begin{figure*}[!t]
  \centering \includegraphics[width=8cm]{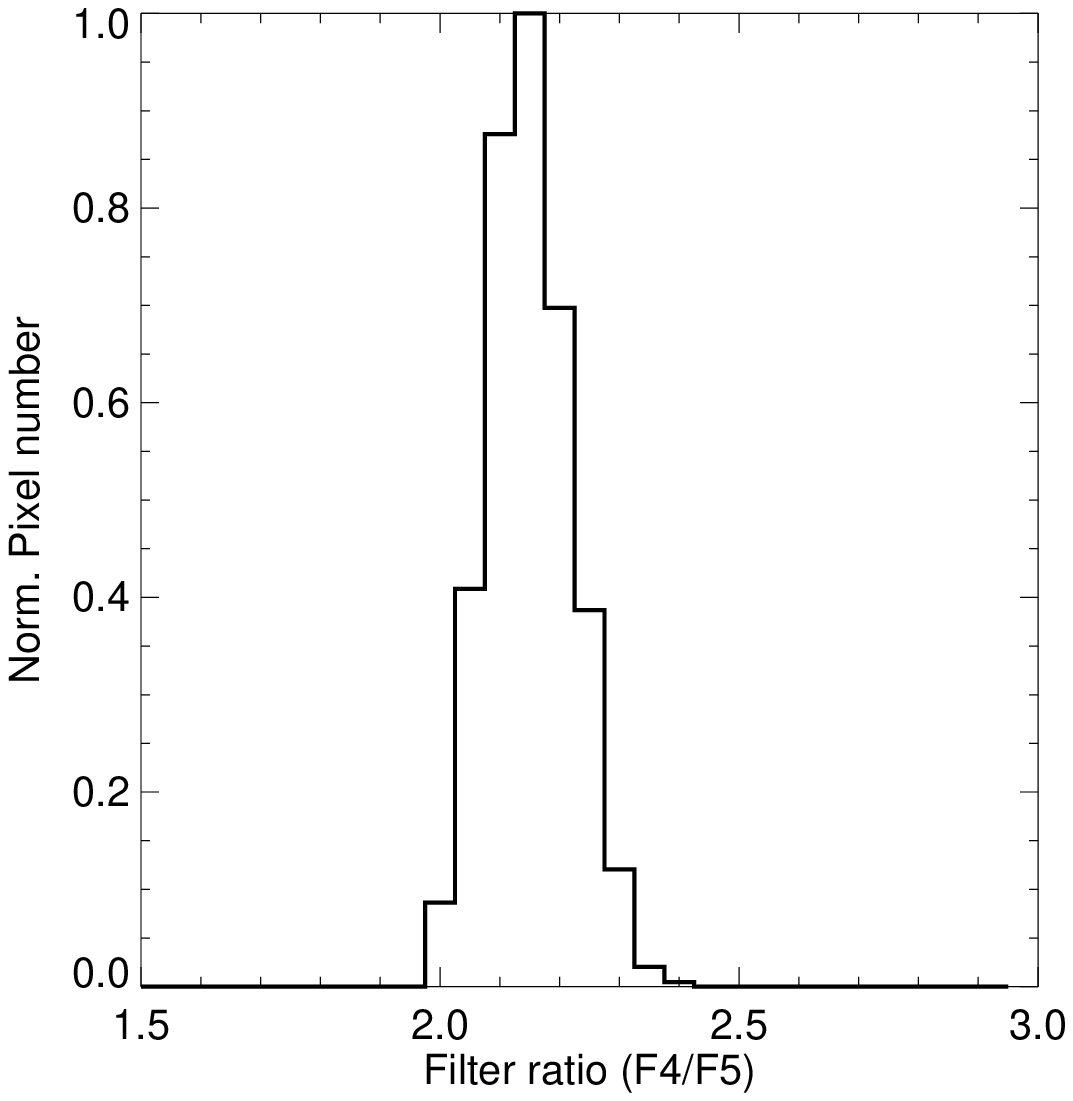} \includegraphics[width=8cm]{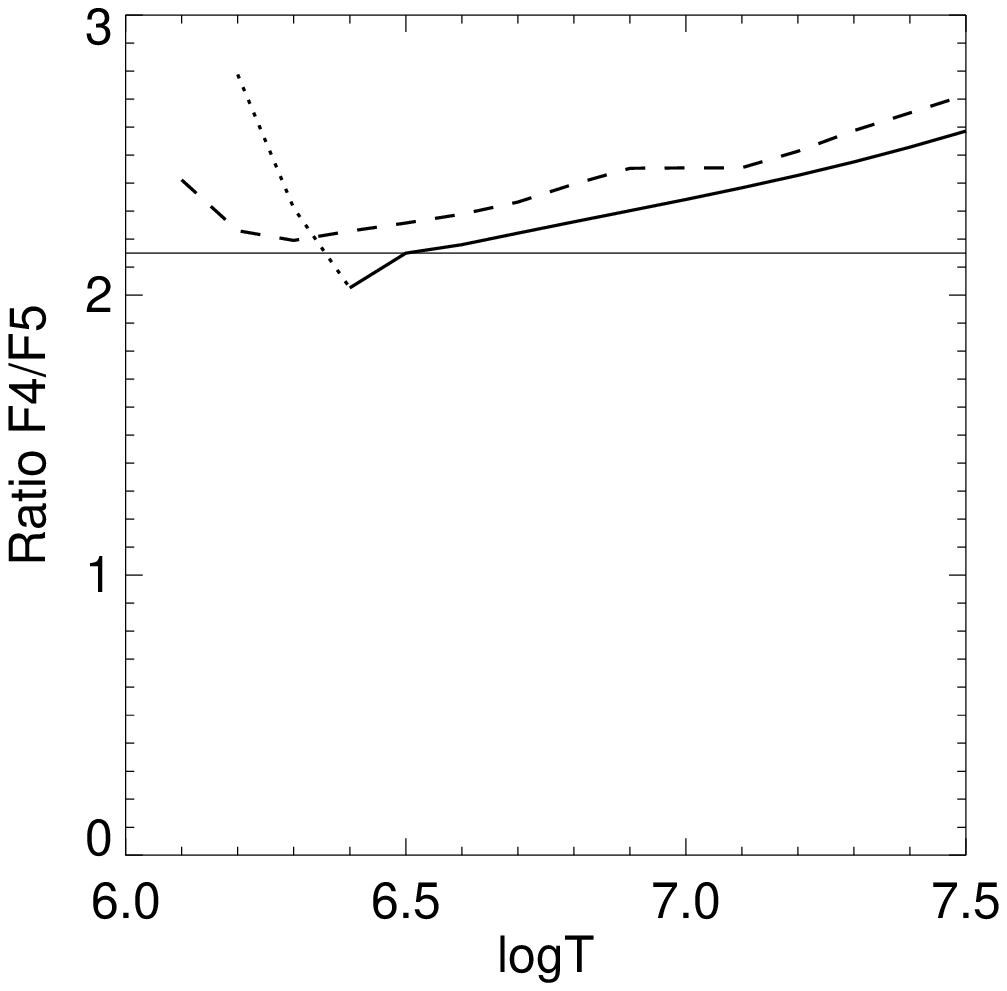}
  \caption{{\it Left:} Distribution of F4/F5 ratio pixel values of the observed active region. The peak is normalized to one. {\it Right:} Dependence of F4/F5 ratio values on temperature (log) as computed with the current release of the filter calibration ({\it dashed line}) and after the fine tuning from feedback with the data, described in this Appendix ({\it solid line}, {\it dotted} below $\log T = 6.4$). The horizontal line marks the peak of the distribution in the left panel.
 }
  \label{fig:app1}
\end{figure*}

By comparing the distributions of measured filter ratio values with
ratios expected from the filter temperature responses for all filter
combinations, we can investigate possible inconsistencies.  On our
data we find that with the new calibration the ratio of the two thickest
filters yields values in a range not entirely compatible with the
expected range, as shown in Fig.~\ref{fig:app1}.
None of the other filter ratios present similar obvious mismatches.
The left panel of Fig.~\ref{fig:app1} shows the distribution of measured F4/F5 filter
ratio values, while the right panel shows the expected filter ratio as a
function of temperature according to the official calibration (dashed line) and after the corrections that we devised here (solid line, the cool branch marked by the dotted line is excluded for the temperature diagnostics); the horizontal line represents the peak
of the observed ratio distribution.  This shows that for a large number of
pixels the measured ratio of hard filters lies outside the range
of values allowed by the current release of the filter calibration, pointing to an obvious problem with the temperature response of one or both involved filters.

We therefore devised a procedure to estimate empirically a correction
factor needed to bring the observations back into agreement with the
expected values and investigated the plausibility of this correction.
We find this correction factor with an iterative method: we
use the distribution of emission measure as derived from the thick filter
analysis as input to synthesize the expected emission in each of the
two filters, and then apply correction factors to the filter temperature
responses in order to reproduce the observed distributions of emission
observed in both filters. 
From the inspection of the changes of the filter responses between the
original and the newly released calibration we find that for most filters the changes in assumed filter thickness roughly correspond to a constant
scaling factor (at least in a range of temperatures where the response
is highest), and therefore we consider a sensible choice
to modify the response by applying a simple scaling factor. At this stage the correction can be applied indifferently to any of the hard filters. Operatively, we chose to act on the F4 filter by 
increasing its response in steps of 5\%. We found that an increase by
about 10\% allows to match the observed values.
We note however that this is the minimum correction factor needed.

\begin{figure*}[!t]
  \centering \includegraphics[width=8cm]{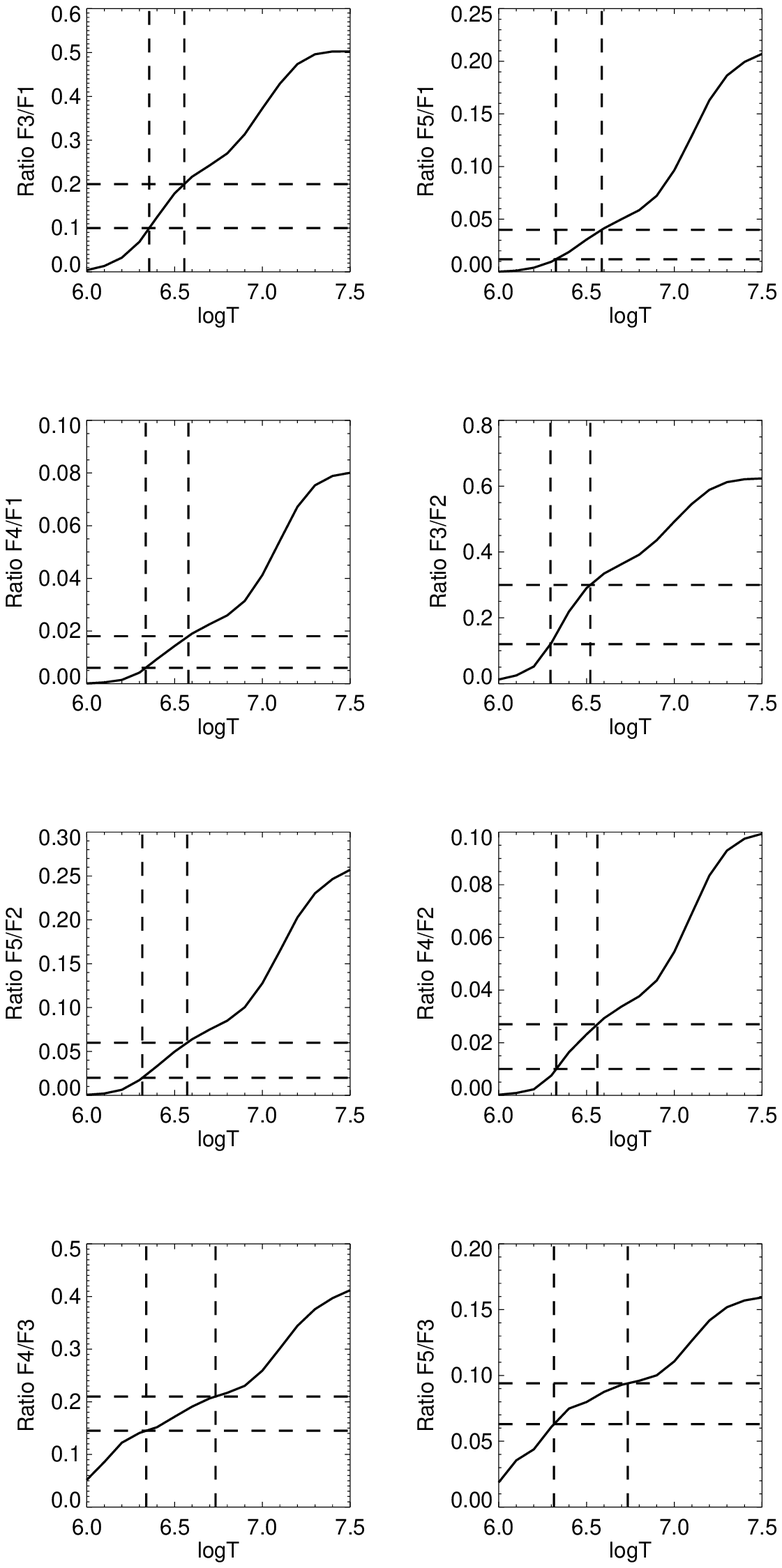} 
  \caption{Dependence of ratio values on temperature (log) as computed after finely tuned calibration for eight different filter ratios. The dashed horizontal lines mark the  ranges of the observed filter ratio distributions (as the one shown in Fig.~\ref{fig:app1}) within 10\% of the peak. The dashed vertical lines bound the corresponding temperature ranges. 
 }
  \label{fig:app2}
\end{figure*}

As further step we examined all other filter ratios to extract additional
information potentially useful to determine which filter is most
likely responsible for the mismatch with observations, and therefore
it is most sensible to correct.  Fig.~\ref{fig:app2} shows the expected filter
ratios for eight filter combinations, using the new calibration after our correction. The horizontal lines in each plot delimit ranges of the observed filter ratio
distributions within 10\% of the peak, and the vertical lines indicate
the corresponding temperature range.  As expected, the ratios of thin
filters give consistent results.  Before correction, the observed ratios of F5/F3
(Al\_med/Be\_thin) corresponded to a wider temperature range with respect to F4/F3
(Be\_med/Be\_thin), in particular covering higher temperatures.  This is not
expected considering that F4 should be relatively more sensitive to
hotter plasma than F5, as indicated by their ratio as a function of
temperature (see Fig.~\ref{fig:app1}). We interpret this as a possible suggestion
that the F5 filter response is the one needing the most
significant correction. We find that the correction factor of $\sim 10$\%
found with the procedure described above yields more consistent results.
Therefore in our analysis we assume an increase of about 10\% to the response
of the F5 filter while the response curves of all other
filters remain as provided by the current XRT calibration.  
We note that this
correction factor is quite reasonable as it corresponds to a variation in
filter thickness of about 3\%, which is compatible with the uncertainties
in filter thickness as derived through laboratory measuments.

Finally, we apply a minor correction to the filter responses
to correct an artifact of the too coarse temperature resolution for
the definition of the temperature responses. As shown in Fig.~\ref{fig:app1} (dashed line) the ratio of the medium filters presents an inflection around $\log T 
\sim 7.0$ due to the insufficient temperature resolution around the peak
of the filter responses. This causes an artificial dip in the derived
distribution of emission measure.  We find that a small correction
($<$2\%) is sufficient to correct for this problem, as shown in Fig.~\ref{fig:app1}.
Further minor corrections are applied in the low temperature range to
obtain further minor improvements. The final correction factor between the current calibration release and our procedure is between 5 and 10\%.

\bibliographystyle{apj}
\bibliography{references}

\end{document}